\documentclass[letterpaper,11pt]{article}
\usepackage{amssymb}
\usepackage{amsmath}
\usepackage{amstext}
\usepackage{graphicx,epsfig}
\usepackage{epsfig}
\usepackage{verbatim} 
\usepackage{fancybox}
\usepackage{color}
\usepackage{ulem}
\usepackage{enumitem}
\usepackage{subfigure}
\usepackage{bbm}
\usepackage{parskip}
\usepackage{dsfont}
\usepackage[numbers,sort&compress]{natbib}
\usepackage[body={6.5in, 9in}, right=1in, top=1in]{geometry}

\newcommand{\Comment}[1]{{}}
\definecolor{darkblue}{rgb}{0.15,0.35,0.55}
\definecolor{reddish}{rgb}{0.65, 0.2, 0.2}
\usepackage[linktocpage=true]{hyperref}
\hypersetup{
colorlinks=true,
citecolor=darkblue,
linkcolor=reddish,
urlcolor=darkblue,
pdfauthor={},
pdftitle={},
pdfsubject={}
}

\newcommand{\be}{\begin{equation}}
\newcommand{\ee}{\end{equation}}
\newcommand{\bea}{\begin{eqnarray}}
\newcommand{\eea}{\end{eqnarray}}
\newcommand{\beas}{\begin{eqnarray*}}
\newcommand{\eeas}{\end{eqnarray*}}

\def\({\left(}
\def\){\right)}

\def\gsim{ \lower .75ex \hbox{$\sim$} \llap{\raise .27ex \hbox{$>$}} }
\def\lsim{ \lower .75ex \hbox{$\sim$} \llap{\raise .27ex \hbox{$<$}} }

\usepackage{xcolor,colortbl}

\begin{document}
\def\thefootnote{\fnsymbol{footnote}}

\begin{center}
\LARGE{\textbf{An Alternative to Particle Dark Matter}} \\[0.5cm]
 
\large{Justin Khoury}
\\[0.5cm]

\vspace{.2cm}

\small{
\textit{Center for Particle Cosmology, Department of Physics and Astronomy, \\ University of Pennsylvania, Philadelphia, PA 19104}}

\vspace{.2cm}

\end{center}

\vspace{.6cm}

\hrule \vspace{0.2cm}
\centerline{\small{\bf Abstract}}
{\small We propose an alternative to particle dark matter that borrows ingredients of MOdified Newtonian Dynamics (MOND) while adding new key components.
The first new feature is a dark matter fluid, in the form of a scalar field with small equation of state and sound speed. This component is critical in reproducing the
success of cold dark matter for the expansion history and the growth of linear perturbations, but does not cluster significantly on non-linear scales. Instead, 
the missing mass problem on non-linear scales is addressed by a modification of the gravitational force law. The force law approximates MOND at large and intermediate accelerations, and therefore reproduces the empirical success of MOND at fitting galactic rotation curves. At ultra-low accelerations, the force law reverts to an inverse-square-law, albeit with a larger Newton's constant.
This latter regime is important in galaxy clusters and is consistent with their observed isothermal profiles, provided the characteristic acceleration scale of MOND is mildly varying with
scale or mass, such that it is $\sim$12 times higher in clusters than in galaxies. We present an explicit relativistic theory in terms of two scalar fields. The first scalar field is governed by a Dirac-Born-Infeld action and behaves as a dark matter fluid on large scales. The second scalar field also has single-derivative interactions and mediates a fifth force that modifies gravity on non-linear scales. Both scalars are coupled to matter via
an effective metric that depends locally on the fields. The form of this effective metric implies the equality of the two scalar gravitational potentials, which ensures  
that lensing and dynamical mass estimates agree. Further work is needed in order to make both the acceleration scale of MOND and the fraction at
which gravity reverts to an inverse-square law explicitly dynamical quantities, varying with scale or mass.} 
\vspace{0.3cm}
\noindent
\hrule
\def\thefootnote{\arabic{footnote}}
\setcounter{footnote}{0}

\section{Introduction}

The Dark Matter (DM) paradigm has been remarkably successful at explaining various large-scale observations. The expansion history, the detailed shape of the peaks in the cosmic microwave background (CMB) anisotropy power spectrum, the growth history of linear perturbations and the shape of the matter power spectrum are all consistent with a non-baryonic, clustering component making up $\sim 25\%$ of the total energy budget. Although this is usually hailed as evidence for weakly interacting {\it particles}, one should keep in mind that these large-scale observations only rely on the {\it hydrodynamical} limit of the dark component. Any perfect fluid with small equation of state ($w\simeq 0$) and sound speed ($c_s \simeq 0$), and with negligible interactions with ordinary matter, would do equally well at fitting cosmological observations on linear scales.

On non-linear scales, the evidence for DM particles is somewhat less convincing. N-body simulations reveal that DM particles self-assemble into halos with a universal density profile, the NFW profile~\cite{Navarro:1996gj}:
\be
\rho_{\rm NFW} (r )= \frac{\rho_s}{\frac{r}{r_s} \left(1 + \frac{r}{r_s}\right)^2} \,.
\ee
The density thus scales as $\sim r^{-1}$ in the interior, and asymptotes to $\sim r^{-3}$ on the outskirts.\footnote{Recent simulations, {\it e.g.}~\cite{Navarro:2008kc}, indicate a shallower slope in the inner regions. The inner profile is closer to the Einasto profiles with $\frac{{\rm d}\ln\rho}{{\rm d}\ln r} \sim -r^{1/n}$, with $n$ slightly varying with halo mass. In particular, $n\simeq 6$ for a Milky-Way size halo, in which case the density profile reverts to $\rho(r) \sim r^{-1}$ at 200~pc from the center.} The regularity of DM self-assembly is certainly a welcome feature. 
Unfortunately, the NFW profile does not naturally account for flat rotation curves of spiral galaxies and the isothermality of galaxy clusters, both of which require $\rho \sim r^{-2}$. The cold dark matter paradigm also faces challenges on small scales, for instance the cuspiness of galactic cores~\cite{Ostriker:2003qj}, the mass~\cite{BoylanKolchin:2011de} and phase-space distributions~\cite{Pawlowski:2012vz,Pawlowski:2013cae,Ibata:2013rh,Ibata:2014csa} of satellite galaxies, and the internal dynamics of tidal dwarfs~\cite{Gentile:2007gp,Kroupa:2012qj,Kroupa:2014ria}. Of course, N-body simulations do not include baryons, so the NFW profile is not expected to hold exactly in the real universe. But the fact that the ``zeroth-order" profile does not readily explain the coarse features of galaxies and clusters of galaxies should at least give us pause. The empirical success or failure of DM particles hinges ultimately on complex baryonic feedback processes.

Quantifying the impact of baryonic physics is an area of active research, but simulations do not yet offer a clear picture. Even qualitative questions, such as whether baryons make the DM profile more cuspy or shallower in the core of galaxies, are still hotly debated~\cite{Ostriker:2003qj}. In the absence of a precise answer, the best one can do when fitting data is incorporate baryonic expectations ({\it e.g.}, adiabatic contraction~\cite{Gnedin:2004cx,Gnedin:2011uj}) through empirical modifications of the NFW profile. Examples include the generalized NFW profile, cored NFW profile, Buckert profile~\cite{Burkert:1995yz}, {\it etc.} See~\cite{Gentile:2004tb} for a recent comparison of how these fare at fitting galactic rotation curves. 

Meanwhile, despite the complexity of baryonic physics, actual structures in our universe show a remarkable level of regularity, embodied in empirical scaling relations. A famous example is the Tully-Fisher relation~\cite{Tully:1977fu}, which relates the luminosity of spiral galaxies to the asymptotic velocity $v_\infty$ of their rotation curves:
\be
L\sim v_\infty^{\,4}\,.
\label{TF}
\ee
Another example is the Faber-Jackson relation~\cite{Faber:1976sn} for elliptical galaxies $L \sim \sigma^4$, where $\sigma$ is the stellar velocity dispersion. These relations are quite puzzling from the particle DM perspective --- why should the rotational velocity in the galactic tail where DM completely dominates be so tightly correlated with the baryonic mass in the inner region? The hope is that these scaling relations will eventually emerge somehow from realistic simulations of coupled baryons and dark matter.

\subsection{MOND empirical law: successes and shortcoming}

MOdified Newtonian Dynamics (MOND) is a radical alternative proposal~\cite{Milgrom:1983ca,Milgrom:1983pn,Milgrom:1983zz}. It attempts to replace dark matter entirely with a modified gravitational force law that kicks in once the acceleration drops to a critical value $a_0$:
\be
a =\left\{\begin{array}{cl} 
a_{\rm N} \qquad ~~~~a_{\rm N} \gg a_0\\[.5cm]
\sqrt{a_{\rm N}a_0} \qquad a_{\rm N} \ll a_0 \,,
\end{array}\right.
\label{MONDlaw}
\ee
where $a_{\rm N} = \frac{G_{\rm N} M(r )}{r^2}$ is the standard Newtonian acceleration. 
By construction, the MOND force law accounts both for the flat rotation curves of spiral galaxies
and the Tully-Fisher relation~\eqref{TF}. Indeed, in the MOND regime the acceleration of a test particle orbiting a spiral galaxy satisfies $\frac{v^2}{r} = \sqrt{\frac{G_{\rm N} Ma_0}{r^2}}$, hence
\be
v^4 = G_{\rm N} Ma_0\,.
\ee
This matches~\eqref{TF} with $M\sim L$. 

\begin{figure}[t]
\centering
\includegraphics[width=5in]{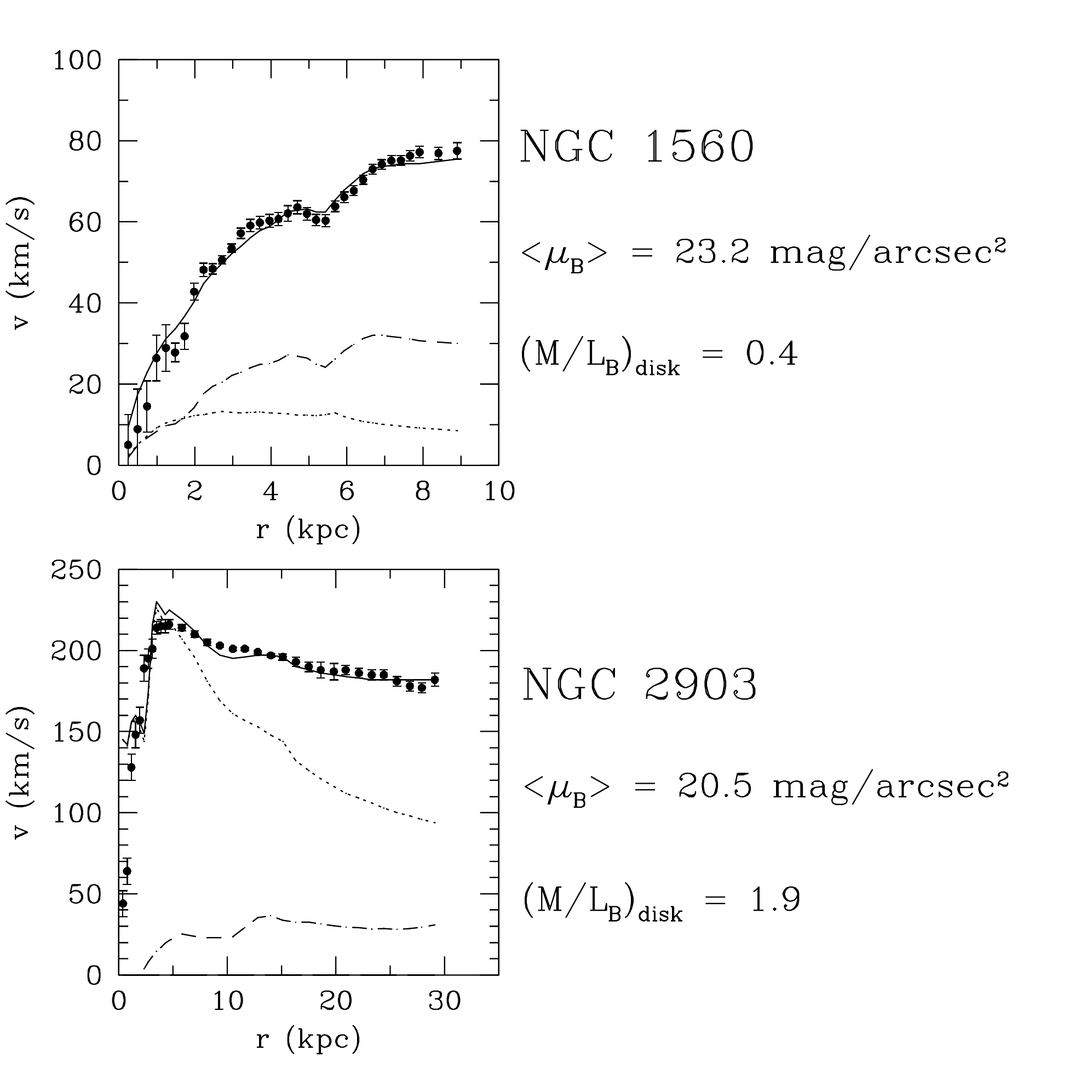}
\caption{\label{rotationcurves} \small Rotations curves from 21~cm observations of LSB galaxy NGC-1560~\cite{Broilsthesis} and HSB galaxy NGC-2903~\cite{Begeman:1991iy}, reproduced from~\cite{Sanders:2002pf}. The dotted and dashed lines are the Newtonian rotation curves from the stellar mass and the gas, respectively. The solid line is the MOND fits, with $a_0$ given by~\eqref{a0bestfitgalaxies}. The only free parameter in each case is the mass-to-light ratio $M/L$.}
\end{figure}

Figure~\ref{rotationcurves}, reproduced from~\cite{Sanders:2002pf}, shows the rotation curves for two galaxies: a low-surface brightness (LSB) galaxy NGC-1560~\cite{Broilsthesis} and the high-surface brightness (HSB) galaxy NGC-2903~\cite{Begeman:1991iy}. The HSB galaxy is in the Newtonian regime within the optical disk and hence approaches the asymptotic velocity with a Keplerian fall-off. The LSB galaxy, on the other hand, is in the MOND regime throughout and hence approaches the asymptotic velocity from below. An intriguing fact is that the best-fit value for the characteristic acceleration is comparable to the Hubble parameter:
\be
a_0^{\rm galaxies} \simeq \frac{1}{6}H_0\simeq 1.2\times 10^{-8}~{\rm cm}/{\rm s}^2\,.
\label{a0bestfitgalaxies}
\ee

The MOND force law has been remarkably successful at explaining a wide range of galactic phenomena, from dwarf galaxies to ellipticals to spirals. See~\cite{Sanders:2002pf,Famaey:2011kh} for comprehensive reviews. It explains the observed upper limit on the surface brightness of spirals, known as Freeman's law~\cite{Freeman:1970mx}, the characteristic surface brightness in ellipticals, known as the Fish law~\cite{fish}, as well as the Faber-Jackson law for ellipticals mentioned earlier. Even if DM particles do exist and gravity is standard, Milgrom's scaling relation~\eqref{MONDlaw} should nonetheless be viewed on the same footing as the Tully-Fisher and Faber-Jackson relations. It is a powerful empirical relation that must be explained by standard theories of galaxy formation. 

\begin{figure}[t]
\centering
\includegraphics[width=5in]{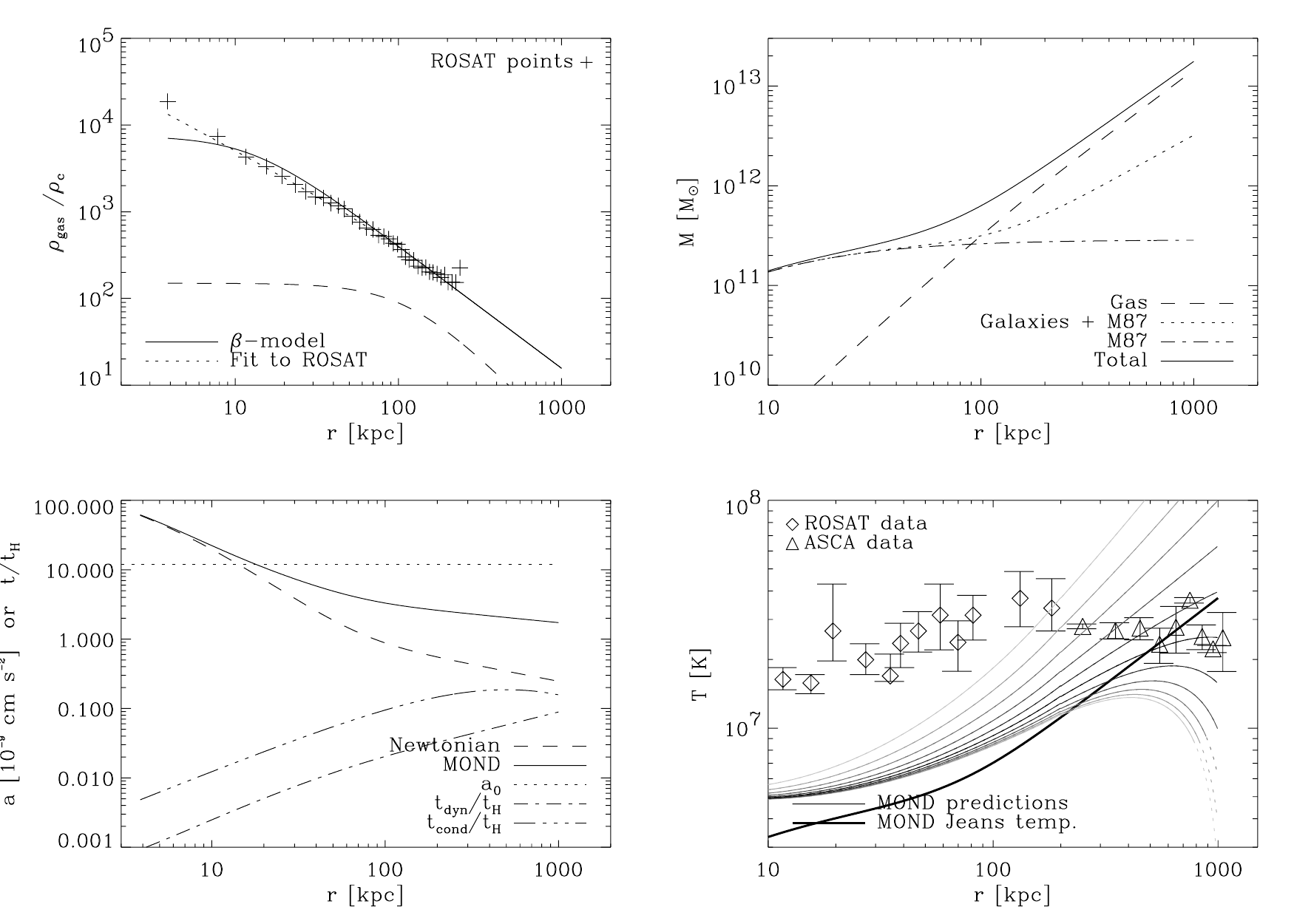}
\caption{\label{Virgo} \small MOND and the Virgo cluster, reproduced from~\cite{Aguirre:2001fj}. The data points are from ROSAT~\cite{nulsen} and ASCA~\cite{Shibata:2000by} observations. The solid lines are the MOND predictions, for different choices of initial temperature at 1~Mpc. The MOND predictions are inconsistent with the nearly isothermal profile.}
\end{figure}

Unfortunately, the empirical success of MOND is limited to galaxies. On cluster scales, the MOND force law fails miserably~\cite{Aguirre:2001fj}. The baryonic component in clusters is dominated by gas, which to a good approximation is in hydrostatic equilibrium and in the MONDian regime. Hydrostatic equilibrium determines the temperature profile $T(r )$ in terms of the observed density profile $\rho(r )$ and the (MONDian) acceleration law $a(r )$. The result does not match the observed isothermal profile of clusters. This is shown in Fig.~\ref{Virgo} for the Virgo cluster, reproduced from~\cite{Aguirre:2001fj}. MOND proponents are forced to assume dark matter, usually in the form of massive neutrinos with $m_\nu\sim 2$~eV~\cite{Sanders:2002ue,Angus:2006ev,Angus:2011hx} and/or cold ($\sim$ 3K), dense gas clouds~\cite{Milgrom:2007cs}.

On cosmological scales, the MOND law requires a relativistic completion. This was achieved just over ten years by Sanders and Bekenstein with a Tensor-Vector-Scalar (TeVeS) theory~\cite{Sanders:1996wk,Bekenstein:2004ne,Sanders:2005vd}. See~\cite{Zlosnik:2006sb} for an elegant reformulation of the theory, and~\cite{Skordis:2008pq,Contaldi:2008iw} for connections to Einstein-aether theories~\cite{Jacobson:2000xp}. (Since TeVeS, other relativistic extensions have been proposed~\cite{Milgrom:2009gv,Blanchet:2009zu,Deffayet:2011sk,Blanchet:2011wv}. See~\cite{Bruneton:2007si} for a review.) First, some good news: perturbations in the vector field accelerate the growth of density perturbations, which allows for the formation of structures. More problematic is the CMB spectrum. An early analysis already revealed some tensions with the height of the third peak~\cite{Skordis:2005xk}, and one would expect that the situation is now much worse with the exquisite data at higher multipoles from the Planck satellite~\cite{Ade:2013zuv} and ground-based experiments~\cite{Sievers:2013ica,Story:2012wx}.\footnote{More precisely, it is possible to get a high third peak without non-baryonic dark matter, but at the cost of distorting the power spectrum on large angular scales~\cite{Zuntz:2010jp}.}
Without a significant dark matter component, the baryonic oscillations in the matter power spectrum tend to be far too pronounced~\cite{Skordis:2005xk,Dodelson:2011qv}. Finally, numerical simulations of MONDian gravity with massive neutrinos fail to reproduce the observed cluster mass function~\cite{Angus:2013sxa,Angus:2014kja}.

\subsection{The best of both worlds}
\label{best}

To summarize, the Cold Dark Matter (CDM) picture is very successful on linear scales, but the jury is still out as to whether it can explain
the detailed structure of galaxies and their empirical scaling relations. MOND, on the other hand, is very successful
on galactic scales, but it seems highly improbable that it can ever be made consistent with the detailed shape of the
CMB and matter power spectra.

In this paper we present a compromise solution: a model which reproduces the CDM phenomenology on linear scales
and reduces to MOND on galactic scales. The model also proposes a key modification to the MOND force law on
cluster scales to explain the observed isothermal profile. In this model, there are {\it no DM particles}. 

The model consists of three key ingredients:

\begin{itemize}

\item To reproduce the CDM phenomenology on large scales, we assume the existence of a perfect fluid 
with small equation of state ($w\simeq 0$) and sound speed ($c_s \simeq 0$). For simplicity, the fluid
is assumed to be irrotational (as vorticity redshifts with the expansion) and barotropic (unique relation between $P$ and $\rho$);
in other words, it is described by a $P(X)$ theory.

This dark component ensures that the cosmology on linear scales is identical to that of the $\Lambda$CDM model. The expansion history,
the linear growth of density perturbations, the detailed shape of the CMB acoustic peaks, and the matter power spectra on scales $\gsim$~Mpc are all
indistinguishable from $\Lambda$CDM predictions.

\item Unlike DM particles, however, the dark fluid does not play a major role on non-linear scales. Instead, the missing mass problem
in galaxies and clusters of galaxy is addressed through a modification to the gravitational force law. In the example of Sec.~\ref{toytheory} below,
this is achieved by a scalar field mediating a fifth force between ordinary matter. The modified force law reduces to MOND on galactic scales,
and therefore reproduces the empirical success of MOND in galaxies. However, the force law deviates from MOND on cluster scales. Specifically,
it approaches an inverse-square law but with a larger Newton's constant.

\end{itemize}

A priori, this hybrid approach to dark matter is not implausible. Any modification to General Relativity (GR) inevitably introduces new degrees of freedom~\cite{Weinberg:1965rz}, and it is
certainly possible that some of these degrees of freedom will act as a dark matter fluid on linear scales. On non-linear scales, however, the new degrees of freedom
modify the gravitational force law. See~\cite{Blanchet:2006yt,Blanchet:2008fj,Zhao:2008rq,Bruneton:2008fk,Li:2009zzh,Ho:2010ca,Ho:2011xc,Ho:2012ar} for other related hybrid proposals.

\section{Le Nouveau MOND}
\label{nouveaumond}

We begin by summarizing the new gravitational force law that reduces to MOND on galactic scales and is modified 
on cluster scales. Unlike MOND, it successfully accounts for the temperature profiles of galaxy clusters. 

\begin{figure}[t]
\centering
\includegraphics[width=6.5in]{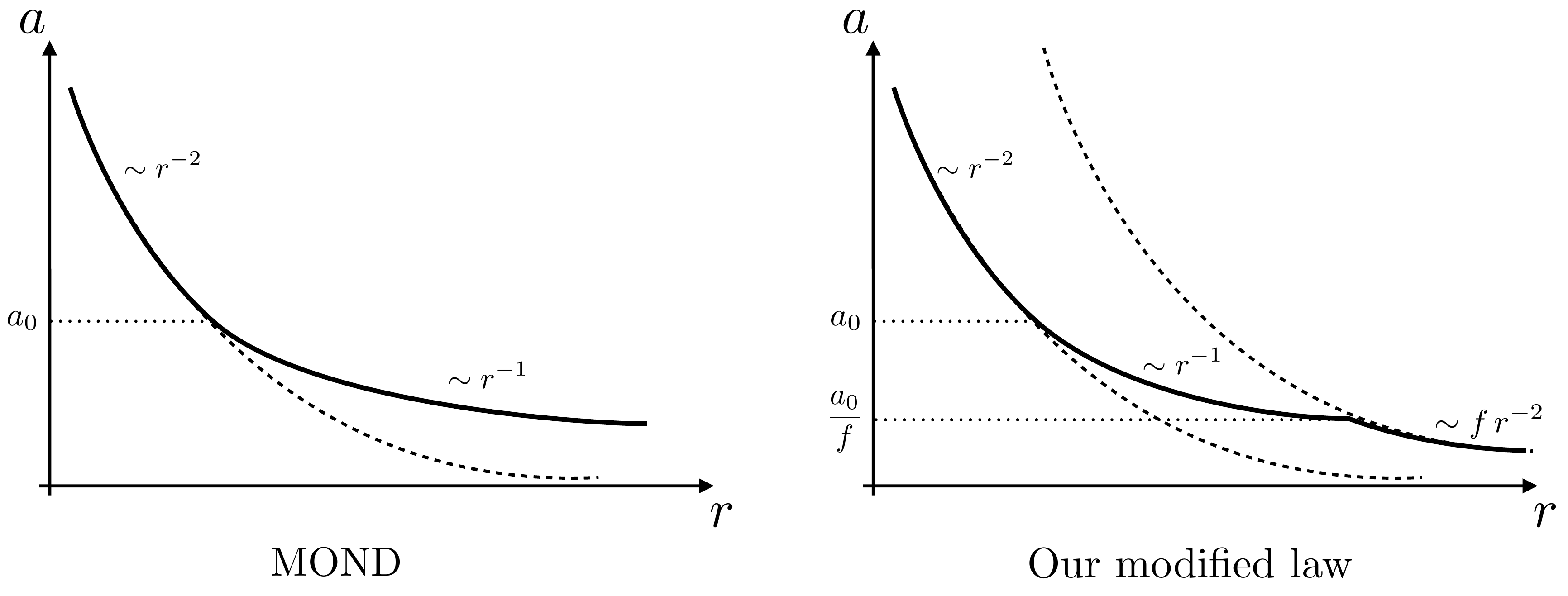}
\caption{\label{forcelaw} \small Sketch of the MOND acceleration law (Left Panel) and our modification to the MOND law (Right Panel), outside a static, spherically-symmetric source. Unlike the MOND case, our modified law reverts back to an inverse-square law at large distances, albeit $f$ times stronger than the standard Newtonian acceleration.}
\end{figure}

The left panel of Fig.~\ref{forcelaw} shows the MOND acceleration for a point mass. In the MONDian regime ($a\ll a_0$), the acceleration is $a \simeq \frac{\sqrt{a_0 G_{\rm N} M}}{r}$. 
Relative to the Newtonian acceleration $a_{\rm N} \simeq \frac{G_{\rm N} M}{r^2}$ (dotted line in the Fig.), the MOND acceleration thus grows without bound:
\be
\frac{a}{a_{\rm N}} \simeq \sqrt{\frac{a_0}{a_{\rm N}}} = \sqrt{\frac{a_0}{G_{\rm N}M}}\,r~\longrightarrow~ \infty\,.
\ee
A related pathology of MOND is that the gravitational energy for a localized source diverges logarithmically.\footnote{The Hamiltonian for the gravitational potential $\Phi$ giving rise to the MONDian acceleration $\vec{a} = -\vec{\nabla}\Phi$ is
\be
H \simeq \int {\rm d}^3x \,a_0^{-1} \left(\vec{\nabla}\Phi\cdot \vec{\nabla}\Phi\right)^{3/2}\,.
\ee
As a result, the gravitational energy for a point charge is divergent:
\be
E = 4\pi\sqrt{a_0 G_{\rm N}^3 M^3} \int^R \frac{{\rm d}r}{r}\sim \log R\,.
\ee
}

Our modified force law instead proposes that this is growth is bounded --- the acceleration eventually reverts back to the inverse-square-law form, but with
a larger Newton's constant: $G_{\rm N}\rightarrow fG_{\rm N}$, where $f > 1$. Specifically, the proposed acceleration is:
\be
a =\left\{\begin{array}{cl} 
a_{\rm N} \qquad ~~~~a_{\rm N} \gg a_0\\[.5cm]
\sqrt{a_{\rm N}a_0} \qquad \frac{a_0}{f^2} \ll a_{\rm N} \ll a_0 \\[.5cm]
fa_{\rm N} \qquad a_{\rm N} \ll \frac{a_0}{f^2} \,.
\end{array}\right.
\label{newMOND}
\ee
This is sketch on the right panel of Fig.~\ref{forcelaw}. From this point of view, the MOND regime is just an interpolation between the ordinary Newtonian acceleration and
a stronger inverse-square-law acceleration. From a boundary-valued standpoint, the recovery of the $1/r$ fall-off behavior for the gravitational potential is a welcome feature:
the gravitational energy for a localized source is finite. A reversal to $1/r^2$ at large distances was first proposed in~\cite{sanders1986} and has since been considered in other contexts~\cite{Moffat:2005si,Famaey:2006iq}. It also been used in~\cite{Bruneton:2007si,Babichev:2011kq} to ensure that the energy of the MONDian field remains positive.

We will see in Sec.~\ref{clusters} that the X-ray temperature profiles of galaxy clusters are well-fitted by the third regime: $a \simeq f a_{\rm N}$.
This relies on a simple yet remarkable fact about clusters: on scales ranging from $\sim 50$~kpc to $\sim 1$~Mpc, the density profile for the gas
is approximately isothermal, $\rho_{\rm gas} \sim 1/r^2$. Not surprisingly, to match the observed temperature the required increase in the strength
of gravity must be comparable to the inferred missing mass: $f \simeq \Omega_{\rm m}/\Omega_{\rm b}\simeq 6$. To ensure that galaxy clusters are
in fact in this third regime, we will find that $a_0$ must be somewhat larger than the value~\eqref{a0bestfitgalaxies} inferred from fitting galaxies, namely
$a_0^{\rm clusters} \simeq 2 H_0\simeq 1.4\times 10^{-7}~{\rm cm}/{\rm s}^2$. This means that $a_0$ must have some mild scale or mass dependence, extrapolating between $\simeq  H_0/6$ on galactic scales to $\simeq  2H_0$ on cluster scales. In Sec.~\ref{galaxies} we will check whether galaxies remain in the MOND regime over the range of scales probed by observations. 
We will find that this is the case if $f$ is somewhat larger on galactic scales, $f\simeq 10$. Hence both $a_0$ and $f$ must be mildly scale or mass dependent.\footnote{The idea that the MOND scale might depend on scale or mass has been pointed out before~\cite{Lue:2003if,Zhao:2012ky}.}

\section{Relativistic Theory: An Example}
\label{toytheory}

The model outlined above allows in principle for various different realizations and variants. To fix ideas, we consider a concrete
example involving two scalar fields. Both are described by ``$P(X)$'' Lagrangians, with single-derivative interactions. We describe
the different ingredients below.

\subsection{Dark Scalar}
\label{DBIscalar}

The first ingredient is a $P(X)$ scalar field:
\be
{\cal L}_\pi = M^4 P(X)\,;\qquad  X \equiv -(\partial\pi)^2 \,.
\label{LDM}
\ee
Thus $\pi$ has dimension of length, and $X$ is dimensionless. 
The stress-energy tensor is
\be
T_{\mu\nu} = M^4 \Big(2P_{,X}\partial_\mu \pi \partial_\nu\pi  + P g_{\mu\nu}\Big)\,.
\label{DMTmunu}
\ee
Identifying a time-like unit vector $u_\mu = \frac{\partial_\mu \pi}{\sqrt{X}}$, this describes 
a perfect, irrotational and barotropic fluid. The equation of state and sound speed are respectively given by~\cite{Garriga:1999vw}
\be
w = \frac{P}{2XP_{,X}-P}\,;\qquad c_s^2 = \frac{P_{,X}}{\rho_{,X}} = \frac{P_{,X}}{2XP_{,XX} + P_{,X}}\,.
\ee
For suitable choice of $P(X)$, both $w$ and $c_s$ can be made small such that the scalar field behaves as dark matter.
For instance, the power-law form $P(X) = X^n$ gives $w = c_s^2 = \frac{1}{2n-1}$, which is small for $n\gg 1$.
Another possibility is to choose $P(X)$ of the ghost condensate form~\cite{ArkaniHamed:2003uy}.
Small perturbations around the ghost condensate redshift as dust and have vanishing $c_s$~\cite{ArkaniHamed:2003uy,Scherrer:2004au,ArkaniHamed:2005gu}.
See~\cite{Peebles:1999fz,Peebles:1999se,Peebles:2000yy,Matos:2000ng,Arbey:2003sj,Guzman:2003kt,Bernal:2006it,Matos:2008ag} for other examples of scalar field dark matter models considered in the literature.

Here we will focus on what is perhaps the most elegant possibility, the Dirac-Born-Infeld (DBI) action~\cite{Silverstein:2003hf}:
\be
{\cal L}_{\rm DBI} = -M_{\rm Pl}^2a_0^2\sqrt{1-X}\,.
\label{LDBI}
\ee
This describes, to lowest-order in derivatives, the motion of a 3-brane in a 5-dimensional space-time. With an eye on the MONDian field discussed below, and to minimize the number of different scales in the theory,
we have set the brane tension to $M^4 \equiv M_{\rm Pl}^2a_0^2$. The induced metric on the brane is
\be
h_{\mu\nu}  = g_{\mu\nu} + \partial_\mu\pi\partial_\nu\pi\,,
\label{hmetric}
\ee
in terms of which ${\cal L}_{\rm DBI} = -M_{\rm Pl}^2a_0^2\sqrt{-h}$. On flat space-time ({\it i.e.}, $g_{\mu\nu} = \eta_{\mu\nu}$), the bulk space-time is Minkowskian,
and the DBI action is protected by the 5d `boost' symmetry,
\be
\delta\pi = v_\mu x^\mu + \pi(x) v^\mu\partial_\mu\pi\,.
\ee
It is straightforward to show that the equation of state and sound speed are given by
\be
w = - \frac{1}{\gamma^2}\,; \qquad c_s = \frac{1}{\gamma}\,,
\ee
where, as usual, $\gamma  \equiv \frac{1}{\sqrt{1-\dot{\pi}^2}}$ is the `Lorentz' factor for the brane motion in the extra dimension. 
Thus the scalar field behaves as dark matter ($w\simeq 0$, $c_s \simeq 0$) in the `relativistic' regime $\gamma \gg 1$, and behaves as
dark energy ($w\simeq -1$, $c_s \simeq 1$) in the `non-relativistic' regime $\gamma \simeq 1$. 

Neglecting the coupling to matter, the background evolution in an expanding FRW universe is governed by
\be
\frac{{\rm d}}{{\rm d}t} \Big( a^3 P_{,X}  \dot{\pi}\Big) = 0\,.
\ee
For DBI, this implies
\be
\frac{\dot{\pi}}{\sqrt{1-\dot{\pi}^2}} = \frac{C}{a^3}\quad \Longrightarrow\quad  \gamma = \sqrt{1 + \frac{C^2}{a^6}}\,,
\ee
where the constant $C$ is determined by initial conditions. Thus $\gamma$ is large in the early universe, decreases as the universe expands, and approaches unity at late times.
We will focus on the situation where $\gamma\gg 1$ up to the present time, {\it i.e.}, $\dot{\pi} \simeq 1$. The case where DBI acts as dark energy today will be
discussed in the Appendix; for reasons explained there, the coupling to matter must be non-local in that case. With $C\gg 1$, the energy density
becomes
\be
\rho = M_{\rm Pl}^2a_0^2 \gamma \simeq M_{\rm Pl}^2a_0^2 \frac{C}{a^3}\,,
\ee
which indeed redshifts like dust. The constant $C$ is fixed by matching to the observed dark matter density today:\footnote{With $a_0 = H_0/6$ and $\Omega_{\rm m} \simeq 0.25$,~\eqref{normC} implies $C = 25$, and therefore $w \simeq C^{-2} \simeq 1.6\times 10^{-3}$ and $c_s \simeq C^{-1} \simeq 0.04$. The fluid thus behaves like dust to an excellent approximation.}
\be
C = 3\Omega_{\rm m} \frac{H_0^2}{a_0^2} \,.
\label{normC}
\ee
In the simplest scenario considered here, this can only be achieved by tuning initial conditions. It would be interesting to study generalizations of the scenario where~\eqref{normC}
would be explained dynamically, for instance by coupling $\pi$ to baryonic matter.

This DBI component reproduces the successful phenomenological success of CDM for the expansion history and linear growth of perturbations. 
Unlike in the standard framework, where DM microscopically consists of weakly-interacting massive particles, here the scalar field is
assumed to be fundamental. The difference appears on non-linear scales. CDM particles cluster to form halos, whereas the DBI fluid does not.
Indeed, as perturbations of $\pi$ grow to become non-linear, the local value of $\gamma = \frac{1}{\sqrt{1 -\dot{\pi}^2 + (\vec{\nabla}\pi)^2}}$ decreases to a value of order unity. At this point, the sound speed
$c_s = \gamma^{-1}$ also becomes order unity, which prevents further clustering. The DBI scalar is therefore protected from developing large gradients and associated caustics.\footnote{This self-protection
is not generic for other choices of $P(X)$. In the ghost condensate example, for instance, caustics may develop --- see~\cite{ArkaniHamed:2005gu} for a detailed discussion. See~\cite{Contaldi:2008iw} for a discussion of caustics in TeVeS.} 

A more quantitative understanding of the $\pi$ profile in the universe clearly requires a careful analysis. A natural expectation is that $\pi$ forms blobs of characteristic size of order the
non-linear scale today ($\sim 1-10$~Mpc). The mass of these blobs would be of order $M_{\rm Pl}^2a_0^2$, which is smaller than the average matter density. Thus $\pi$ should give a
small correction to the typical mass fluctuation on non-linear scales.\footnote{We thank Paolo Creminelli for helpful discussions on this point.} For the purpose of this paper, we will ignore the
spatial gradients of $\pi$ and treat it as a homogeneous component. On non-linear scales, the `missing mass' problem is instead solved by a second scalar field $\phi$ which modifies the gravitational force law.

\subsection{Extending MOND}
\label{newforce}

The second ingredient is another derivatively-coupled scalar field:
\be
{\cal L}_{\rm New\,MOND} = M_{\rm Pl}^2a_0^2 \, F(Y)\,;\qquad  Y \equiv - \frac{(\partial\phi)^2}{M_{\rm Pl}^2a_0^2} \,.
\label{Lnewmond}
\ee
Unlike $\pi$, which has mass dimension $-1$, $\phi$ has the standard mass dimension $+1$. This field has negligible impact on the background evolution. Its role is limited to
modifying the gravitational force law between ordinary matter sources. The exact form of the matter coupling will be discussed below, but for the moment let us focus on
non-relativistic matter and assume the coupling
\be
{\cal L}_{\rm coupling} \simeq -\frac{\phi}{M_{\rm Pl}} \rho\,. 
\label{NRcoup}
\ee
In the quasi-static approximation, the equation of motion for $\phi$ reduces to
\be
\vec{\nabla}\cdot \Big( F_{,Y} \vec{\nabla}\phi \Big) = \frac{\rho}{2M_{\rm Pl}}\,.
\label{Newtonphi}
\ee
The total acceleration on a test particle is
\be
\vec{a} = -\vec{\nabla}\left( \Phi_{\rm N} + \frac{\phi}{M_{\rm Pl}}\right) = \vec{a}_{\rm N}  -\frac{\vec{\nabla}\phi}{M_{\rm Pl}} \,.
\label{totalacc}
\ee

To reproduce the modified acceleration law~\eqref{newMOND}, we claim the function $F$ must satisfy\footnote{Note that
$Y \simeq  - \frac{(\vec{\nabla}\phi)^2}{M_{\rm Pl}^2a_0^2} < 0$ in the quasi-static approximation.}
\bea
F(Y) \simeq \left\{\begin{array}{cl} 
-\frac{2}{3} (-Y)^{3/2} \qquad~ & |Y|  \gg \frac{1}{f^2}  \\[.5cm]
\frac{Y}{f} \qquad ~&  |Y|  \ll  \frac{1}{f^2} \,.
\end{array}\right.
\label{newF}
\eea
Let us check the two regimes in turn. For simplicity, we will assume $f\gg 1$.

\begin{itemize}

\item {\bf ``MONDian'' regime:} In the first regime, where $F(Y) \simeq -\frac{2}{3} (-Y)^{3/2}$, the equation of motion~\eqref{Newtonphi} reduces to
\be
\vec{\nabla} \cdot \left( \frac{| \vec{\nabla}\phi |}{M_{\rm Pl} a_0}  \vec{\nabla}\phi\right) = \frac{\rho}{2M_{\rm Pl}}\,.
\ee
For a static, spherically-symmetric source, this integrates to
\be
\frac{\phi'}{M_{\rm Pl}} = \sqrt{a_0 \frac{G_{\rm N}M(r )}{r^2}} =  \sqrt{a_0 a_{\rm N}}\,,
\ee
which is the MONDian form. However, whether the {\it total} acceleration~\eqref{totalacc} is approximately Newtonian or MONDian depends on
whether $\Phi'_{\rm N} \gg \phi'/M_{\rm Pl}$ or $\ll \phi'/M_{\rm Pl}$. In other words, the acceleration is Newtonian ($a\simeq a_{\rm N}$) whenever $a_{\rm N} \gg a_0$,
and approximately MONDian ($a \simeq  \sqrt{a_0 a_{\rm N}}$) whenever $a_{\rm N} \ll a_0$, as desired.

\item {\bf Inverse-square-law regime:} In the second regime, where $F(Y) \simeq Y/f$, we instead have
\be
\vec{\nabla}^2\phi = f \frac{\rho}{2M_{\rm Pl}}\,,
\ee
which can be rearranged as
\be
\vec{\nabla}^2 \frac{\phi}{M_{\rm Pl}} = 4\pi f G_{\rm N} \rho\,.
\ee
This is identical to Poisson's equation for the gravitational potential, with Newton's constant rescaled by a factor of $f$. In the limit $f\gg 1$, the acceleration is dominated by
$\phi$-exchange and given by
\be
\vec{a} \simeq f \vec{a}_{\rm N}\,.
\ee 

\end{itemize}

\noindent The transition between two regimes occurs when $\frac{|\vec{\nabla} \phi|}{M_{\rm Pl}} \sim \sqrt{a_0 a_{\rm N}} \sim fa_{\rm N}$, {\it i.e.}, when 
\be
\frac{|\vec{\nabla} \phi|}{M_{\rm Pl}} \sim \frac{a_0}{f}\,.
\ee
In other words, the transition occurs when $Y \sim 1/f^2$, as claimed in~\eqref{newF}.

It is well-known that a MONDian fifth force $\phi' \sim \sqrt{a_0 a_{\rm N}}$ gives too large a correction to Newtonian gravity in the solar system to
be consistent with local tests of gravity. One possible way out is to suitably modify $F(Y)$ at large $Y$, but this requires fine-tuning~\cite{Bruneton:2007si}. A much more
elegant solution was proposed recently based on galileons and Vainshtein screening~\cite{Babichev:2011kq}. One simply adds to the action the galileon
operator~\cite{Deffayet:2009wt,Deffayet:2009mn,deRham:2010eu,VanAcoleyen:2011mj}
\be
{\cal L}_{\rm Galileon} = - \frac{\ell^4}{3} \varepsilon^{\alpha\beta\gamma\delta} \varepsilon^{\mu\nu\rho\sigma} R_{\gamma\delta\rho\sigma} \partial_\alpha\phi \partial_\mu\phi\nabla_\nu\nabla_\beta \phi\,,
\ee
where $\ell$ has units of length, and $\varepsilon^{\mu\nu\rho\sigma}$ is the Levi-Civita tensor. This operator introduces a new scale, the Vainshtein scale $r_{\rm V} \sim \left(G_{\rm N} M a_0\right)^{1/4}\ell$, below which
the scalar profile is modified. This can restore consistency with solar system tests if $\ell \; \lsim\; 100$~kpc~\cite{Babichev:2011kq}. On scales larger than $r_{\rm V}$, this new operator is negligible.

An important consideration is the stability of perturbations. Expanding~\eqref{Lnewmond} around a spherically-symmetric background, $\phi = \bar{\phi}(r) + \varphi$,
we find at quadratic order
\be
{\cal L}_{\rm quad} = F_{,Y} \left(\dot{\varphi}^2 - (\partial_\Omega\varphi)^2\right) - \Big(F_{,Y} + 2 Y F_{,YY}\Big)\varphi'^2 \,.
\label{quadaction}
\ee
To avoid ghosts, we clearly need $F_{,Y} > 0$. This is satisfied in both regimes of~\eqref{newF}. To avoid gradient instabilities in the radial direction, we also need
$F_{,Y} + 2 Y F_{,YY} > 0$, which is also satisfied by~\eqref{newF}. However, the sound speed of radial propagation,
\be
c_s^{\rm radial} = \sqrt{1+  \frac{2 Y F_{,YY}}{F_{,Y}}}\,,
\ee
is strictly superluminal. Indeed, in the MONDian regime where $F(Y) \sim (-Y)^{3/2}$, we have $c_s^{\rm radial}\simeq \sqrt{2}$. 
This fact, first observed long ago~\cite{Bekenstein:1984tv}, is not surprising: our $F(Y)$ theory is an example of the kinetic or k-mouflage screening mechanism~\cite{Babichev:2009ee,Brax:2012jr,Burrage:2014uwa}. (See~\cite{Jain:2010ka,Joyce:2014kja} for reviews of screening mechanisms and observational tests.) Indeed, the scalar force is much smaller than the Newtonian force in the limit $Y \rightarrow -\infty$, that is, it is screened. It is well-known that derivative screening comes hand-in-hand with superluminality~\cite{Dvali:2010jz}. In particular, the UV completion of the theory cannot be
a local quantum field theory~\cite{Adams:2006sv}. It has been conjectured in certain examples that chronology protection may prevent the formation of closed causal curves~\cite{Babichev:2007dw,Burrage:2011cr,Burrage:2014uwa}, in analogy with Hawking's Chronology Protection Conjecture in GR~\cite{Hawking:1991nk}. At a more basic level, whether a theory truly exhibits superluminal propagation
can be somewhat ambiguous at the effective field theory level~\cite{deRham:2014lqa}.

Obviously there are many different choices of $F$ that are consistent with~\eqref{newF}. In particular,~\eqref{newF} only constrains
the functional form for $Y < 0$; the region $Y > 0$, relevant for the cosmological evolution and linear perturbations, is
completely unconstrained. Note that, around time-dependent backgrounds, the quadratic action for perturbations takes a form similar to~\eqref{quadaction}, with the time and
radial components interchanged. Thus the conditions for absence of ghosts and gradient instabilities are the same: 
\bea
\nonumber
F_{,Y} &>& 0\,;\\
F_{,Y} + 2 Y F_{,YY} &>& 0\,. 
\label{Fcond}
\eea
These conditions ensure that the Hamiltonian is bounded below and the Cauchy problem is well-defined~\cite{Bruneton:2007si,Babichev:2007dw}.

As a concrete example, consider the ``DBI-like'' form
\be
F(Y)  = \frac{Y}{f} \sqrt{ 1 - \left(\frac{2f}{3}\right)^2 Y}\,,
\label{DBIlike}
\ee
which clearly has the desired limits~\eqref{newF}. It also satisfies~\eqref{Fcond} for all $Y < 0$, but fails to do so for $ Y \; \gsim\;  \frac{3}{5f^2}$. Another
option which satisfies~\eqref{Fcond} for all $Y$ is
\be
F(Y)  = \frac{Y}{f} \left( 1+ \left(\frac{2f}{3}\right)^4 Y^2\right)^{1/4}\,.
\label{Feg}
\ee

Another desirable property of $F$ is that, in the presence of gravity, it admits a positive energy theorem for asymptotically flat solutions~\cite{Schon:1979rg,Witten:1981mf,Parker:1981uy,Nester:1982tr}. That is, 
the ADM mass should be non-negative, vanishing only for the trivial Minkowski solution. Recently, the standard arguments for canonical scalar fields~\cite{Boucher:1984yx,Townsend:1984iu,Deser:2006gt}
have been generalized to $P(X,\phi)$ theories~\cite{Elder:2014fea}. See also~\cite{Nozawa:2013maa}. A sufficient (but not necessary) condition to have positive energy is if $Y F_{,Y} - F $ is bounded from below. This is not the case for the example~\eqref{Feg}: it is easily seen that $Y F_{,Y} - F $ is negative definite and $\rightarrow -\infty$ as $Y \rightarrow -\infty$. To guarantee positive energy, one could either
modify $F$ at large $|Y|$, or restrict the range of allowed $Y$. 

\subsection{Coupling to matter}

The third ingredient is the coupling of $\phi$ and $\pi$ to matter fields. Inspired by TeVeS~\cite{Bekenstein:2004ne}, we assume that the matter action is of
the form $S_{\rm m}[\tilde{g}_{\mu\nu}, \psi]$, with matter fields coupling to the metric
\bea
\nonumber
\tilde{g}_{\mu\nu} &=& e^{-2\phi/M_{\rm Pl}} h_{\mu\nu} - e^{2\phi/M_{\rm Pl}} \partial_\mu\pi \partial_\nu\pi \\
&=& e^{-2\phi/M_{\rm Pl}} g_{\mu\nu} - 2\partial_\mu\pi \partial_\nu\pi \sinh \frac{2\phi}{M_{\rm Pl}} \,.
\label{gmat}
\eea
where $h_{\mu\nu}$ was defined in~\eqref{hmetric}. Note that this metric is a {\it local} function of the fields, unlike other forms considered in the literature, {\it e.g.}~\cite{Blanchet:2011wv}.
The metric $\tilde{g}_{\mu\nu}$ is invariant under
\bea
\nonumber 
h_{\mu\nu} &\rightarrow &  e^{2\lambda} h_{\mu\nu}  \\
\nonumber
\pi &\rightarrow & e^{-\lambda}\pi   \\
\phi  & \rightarrow & \phi + \lambda M_{\rm Pl}\,.
\eea
This can be promoted to a symmetry of the full theory (except for the Einstein-Hilbert term) by replacing
$g_{\mu\nu}$ with $\tilde{g}_{\mu\nu}$ in the $\pi$ and $\phi$ actions. In other words, in~\eqref{LDM} and~\eqref{Lnewmond}
we would make the replacements
\bea
X = -(\partial\pi)^2 &\rightarrow & - \tilde{g}^{\mu\nu}\partial_\mu\pi\partial_\nu\pi \,;\\
Y = - \frac{1}{M_{\rm Pl}^2a_0^2}  (\partial\phi)^2 &\rightarrow & - \frac{1}{M_{\rm Pl}^2a_0^2} \tilde{g}^{\mu\nu}\partial_\mu\phi\partial_\nu\phi\,.
\eea
Since $\tilde{g}_{\mu\nu} \simeq  g_{\mu\nu}$ to leading order in $\phi/M_{\rm Pl}$, this substitution has negligible effect
on the dynamics of $\phi$ and $\pi$ described earlier.

The form of the metric~\eqref{gmat} is critical to get the correct lensing signal. In the weak-field, quasi-static regime, the Einstein-frame metric
takes the usual form:
\be
g_{\mu\nu} {\rm d}x^\mu {\rm d}x^\nu = -(1 + 2\Phi){\rm d}t^2 + (1-2\Phi){\rm d}\vec{x}^2\,.
\label{GRform}
\ee
Furthermore, assuming that the $\pi$ profile is not dramatically altered by the presence of the source, we still have $\dot{\pi} \simeq 1$ locally.
For small $\phi$, we can therefore approximate $\partial_\mu\pi \partial_\nu \pi \sinh \frac{2\phi}{M_{\rm Pl}}\simeq \frac{2\phi}{M_{\rm Pl}} \delta_\mu^{\;0}\delta_\nu^{\;0}$,
yielding the effective metric:
\be
{\rm d}\tilde{s}^2 = \tilde{g}_{\mu\nu} {\rm d}x^\mu {\rm d}x^\nu = -\left(1 + 2\left[\Phi + \frac{\phi}{M_{\rm Pl}}\right] \right){\rm d}t^2 + \left(1 - 2\left[\Phi + \frac{\phi}{M_{\rm Pl}}\right] \right){\rm d}\vec{x}^2\,.
\label{lensingform}
\ee
This is exactly of the GR form~\eqref{GRform}, albeit in terms of a shifted gravitational potential
\be
\tilde{\Phi} = \Phi + \frac{\phi}{M_{\rm Pl}}\,.
\ee
In particular, the mass inferred from lensing observations precisely matches the mass inferred from dynamical measurements, as desired.\footnote{As a check on our earlier results,
the $\phi$ coupling to a quasi-static source $T^{\mu\nu} \simeq \rho \delta^\mu_{\;0}\delta^\nu_{\;0}$ is
\be
{\cal L}_{\rm coupling} =  \frac{1}{2} \tilde{g}_{\mu\nu} \rho \delta^\mu_{\;0}\delta^\nu_{\;0} \supset - \frac{\phi}{M_{\rm Pl}}\rho\,,
\ee
which is consistent with~\eqref{NRcoup} assumed earlier.}

Notice that the DM fluid plays a dual role in our scenario: $i)$ it acts as dark matter on large scales to reproduce the $\Lambda$CDM phenomenology for the expansion history and linear growth; 
$ii)$ it offers, through the scalar field time-derivative, an effective ``aether'' for the coupling to matter, which is essential for lensing.
 
\begin{table}[float]
\centerline{
\small
\begin{tabular}{| l | c | c | }\hline 
\raisebox{8pt} {\phantom{M}}~~~~~~~{\bf Scale} \raisebox{-8pt}{\phantom{M}} & \raisebox{8pt} {\phantom{M}}$a_0$  \raisebox{-8pt}{\phantom{M}}  & \raisebox{8pt} {\phantom{M}} $f$   \raisebox{-8pt}{\phantom{M}} \\ \hline
\raisebox{8pt} {\phantom{M}}	Galactic (Sec.~\ref{galaxies}) \raisebox{-8pt}{\phantom{M}}  & \raisebox{8pt} {\phantom{M}}$\simeq \frac{1}{6}H_0$  \raisebox{-8pt}{\phantom{M}} & \raisebox{8pt} {\phantom{M}}$\gsim \; 10$  \raisebox{-8pt}{\phantom{M}} \\		\hline
\raisebox{8pt} {\phantom{M}}	Cluster (Sec.~\ref{clusters})  \raisebox{-8pt}{\phantom{M}} & \raisebox{8pt} {\phantom{M}}$\gsim\; 2H_0$  \raisebox{-8pt}{\phantom{M}} & \raisebox{8pt} {\phantom{M}}$\simeq 6$  \raisebox{-8pt}{\phantom{M}}  \\				\hline
\raisebox{8pt} {\phantom{M}}	Cosmological  \raisebox{-8pt}{\phantom{M}} & \raisebox{8pt} {\phantom{M}} $\sim H_0$~?  \raisebox{-8pt}{\phantom{M}} &\raisebox{8pt} {\phantom{M}} $\lesssim 1$~?  \raisebox{-8pt}{\phantom{M}} \\ 							\hline
\end{tabular}
}
\caption{\small Constraints on the parameters $a_0$ and $f$ of the modified force law on different scales.}
\label{params}
\end{table}

\subsection{Scale dependence}
\label{scaledep}

As we will see in great detail in the following Sections, the parameters $a_0$ and $f$ of the modified force law must vary mildly with scale or mass in order to simultaneously reproduce the phenomenology of galaxies and clusters of galaxies. 
The required values of $a_0$ and $f$ on different scales are summarized in Table~\ref{params}.

\begin{itemize}

\item On galactic scales, the parameters are constrained by demanding that the successful MONDian phenomenology is reproduced, from dwarf galaxies to large spiral galaxies.
In particular, $a_0$ must assume the preferred MOND value of $\simeq H_0/6$. Meanwhile, $f$ must be large enough to ensure that the MONDian regime applies to the smallest galaxies.
In Sec.~\ref{galaxies}, we will find this is the case for $f\; \gsim\;  10$. 

\item On cluster scales, the constraint on $a_0$ comes from demanding that clusters are in the enhanced inverse-square-law regime, instead of the MOND regime. We will find in Sec.~\ref{clusters}
that this requires $a_0\; \gsim\; 2H_0$. Meanwhile, the value of $f\simeq 6$ is set by normalizing to the observed X-ray temperatures. 

\item On cosmological scales, the constraints are not as stringent. The value of $a_0$ is relatively unconstrained, though obviously the most natural possibility is $a_0\sim H_0$. 
The value of $f$, however, must be somewhat smaller than for clusters, {\it e.g.}, $f\lesssim 1$. If the value of $f$ is too large, then the scalar-mediated force will lead to
an unacceptably large growth rate of density perturbations.

\end{itemize}

\noindent The required scale (or mass) dependence is fairly mild --- a logarithmic behavior would suffice. This has clearly not been included in the relativistic example described so far, where $a_0$ and $f$ have been treated as constants. This clearly points towards making $a_0$ and $f$ dynamical. The most elegant possibility would be to explain the appearance of $H_0$ in the MONDian Lagrangian through a dynamical mechanism. If $a_0$ and $f$ are determined cosmologically, then we can expect them to vary with scale as well. One possibility to achieve the desired scale/mass dependence of these parameters is if their value depends on an environmentally-dependent scalar field, such as in the chameleon~\cite{Khoury:2003aq,Khoury:2003rn,Brax:2004qh,Gubser:2004uf} or symmetron mechanisms~\cite{Hinterbichler:2010es,Hinterbichler:2011ca}. We leave this to future work and for now turn our attention to observations.

\section{Galaxy Clusters}
\label{clusters}

In this Section we look more closely at galaxy clusters to justify the modification to the MOND force law proposed in Sec.~\ref{nouveaumond}. As is well-known, galaxy clusters are dominated
by baryonic gas (and dark matter, in the conventional picture), known as the intra-cluster medium (ICM). Assuming spherical symmetry and hydrostatic
equilibrium, for simplicity, the density and pressure of the gas are related to the gravitational acceleration by
\be
\frac{1}{\rho} \frac{{\rm d}P}{{\rm d} r} = -a \,.
\label{hydroequi}
\ee
Approximating the gas as ideal, then
\be
P = \frac{\rho}{\mu m_{\rm p}}kT\,,
\label{eos}
\ee
where $m_{\rm p} = 938$~MeV is the proton mass, and $\mu \simeq 0.59$ is the mean molecular weight per particle for a fully ionized plasma
with hydrogen mass fraction $1 - Y = 0.76$. Combining~\eqref{hydroequi} and~\eqref{eos}, we obtain a differential equation relating the density and temperature profiles to the acceleration:
\be
\frac{{\rm d}\ln \rho}{{\rm d}\ln r} + \frac{{\rm d}\ln T}{{\rm d}\ln r} = - \frac{\mu m_{\rm p}}{kT} r\,a \,.
\label{hydroequiODE}
\ee

\begin{figure}
\centering
\includegraphics[width=6.5in]{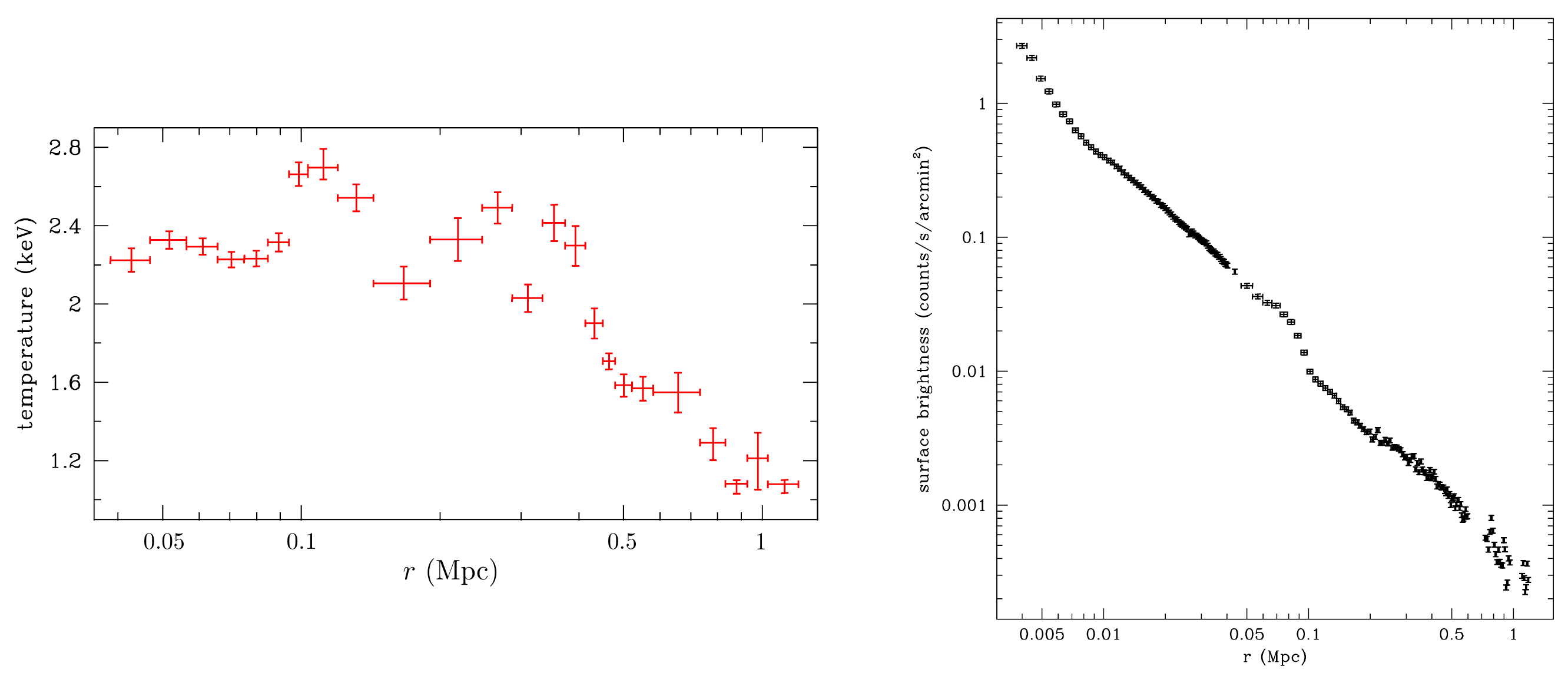}
\caption{\label{VirgoXMM} \small Temperature (Left Panel) and surface brightness (Right Panel) profiles for the Virgo cluster from the XMM-Newton satellite, reproduced from~\cite{Urban:2011ij}.}
\end{figure}

X-ray observations measure the ICM density and temperature up to distances $\lesssim$~Mpc from the center. (We will discuss shortly Sunayev-Zeldovich (SZ) and weak lensing observations which probe larger distances.) This is illustrated in Fig.~\ref{VirgoXMM} again for the Virgo cluster, but this time with more recent data from the XMM-Newton satellite~\cite{Urban:2011ij}. The left panel shows the (projected) temperature profile. The right panel shows the radial surface brightness. To zeroth approximation, the ICM temperature is constant over the range of scales probed, $50~{\rm kpc} \lesssim r \lesssim 1~{\rm Mpc}$, {\it i.e.},
\be
\frac{{\rm d}\ln T}{{\rm d}\ln r} \approx 0\,.
\ee
The surface brightness is generally well-fitted by the $\beta$-model~\cite{Cavaliere:1976tx}, with $I(r ) \sim r^{-6\beta + 1}$ outside the central region. 
The corresponding (deprojected) radial density is $\rho(r )\sim r^{-3\beta}$. The value of $\beta$ varies from cluster to cluster, of course,
but a typical value is $\beta \approx 2/3$, corresponding to $I(r ) \sim r^{-3}$ and $\rho(r )\sim r^{-2}$. This is the {\it isothermal} profile.
A quick look at the right panel of Fig.~\ref{VirgoXMM} shows that $I(r ) \sim r^{-3}$ is indeed a good approximation for $r \,\gsim\, 10~{\rm kpc}$. Therefore,
\be
\frac{{\rm d}\ln \rho}{{\rm d}\ln r} \approx -2\,.
\ee
With $T(r ) \simeq {\rm const.}$ and $\rho(r )\sim r^{-2}$,~\eqref{hydroequiODE} requires $a(r )\sim r^{-1}$. This is satisfied
for an inverse-square law $a(r ) \sim M(r )/r^2$, which is of course why the density profile $\rho(r )\sim r^{-2}$ is called isothermal. 

Suppose that, over the relevant scales, clusters are in the enhanced inverse-square-law regime:
\be
a(r ) = f\frac{G_{\rm N}M(r )}{r^2}\,.
\label{fNewton}
\ee
With $\rho\sim r^{-2}$, the mass enclosed within a given radius is
\be
M_{\rm ICM}(r ) = 4\pi \int_0^r {\rm d}r' r'^{\,2} \rho(r')  \sim r\,,
\ee
which implies $a(r )\sim r^{-1}$, as claimed. The constant $f$ can be fixed by normalizing to the observed temperature. Mohr {\it et al.}~\cite{Mohr:1999ya} studied a sample of 45 galaxy clusters using the ROSAT X-ray data. At fixed radius $r = 0.7$~Mpc,\footnote{We assume $H_0 = 70~{\rm km} {\rm s}^{-1} {\rm Mpc}^{-1}$ to convert the distance scales quoted in~\cite{Mohr:1999ya}. It is worth stressing that the quoted mass is at {\it fixed radius}, as opposed to the virial radius, which explains why the value may at first sight appear smaller than expected.} they obtained the following mass-temperature relation
\be
M_{0.7\;{\rm Mpc}} = \left(0.82 \pm 0.05\right)\times 10^{13}M_\odot \left(\frac{kT}{{\rm keV}}\right)^{1.23\pm 0.17}\,.
\label{mohrfit}
\ee
Assuming an isothermal profile and the force law~\eqref{fNewton}, we obtain
\be
M_{0.7\;{\rm Mpc}} = 0.86 \times 10^{13}M_\odot \cdot \frac{kT}{{\rm keV}} \cdot \frac{6}{f} \,.
\ee
The linear dependence on $T$ is consistent with~\eqref{mohrfit} within error bars. We obtain a remarkably good fit to the normalization for
\be
f_{\rm clusters} = 6\,.
\ee

The mass-temperature relation for clusters is often expressed in terms of the mass at the virial radius. In the standard CDM picture, the
spherical collapse model famously predicts~\cite{mo}
\be
M_{200} \sim T^{3/2}\,,
\label{MT}
\ee
where $M_{200}$ is the mass when the cluster density reaches 200 times the critical density, which is when virialization should occur.\footnote{Note that $M_{200}$ is different
than the mass at a fixed physical radius, $M_{0.7\;{\rm Mpc}}$. In the standard CDM picture, these are related through the density profile, usually assumed to be NFW. Therefore,
there is no contradiction {\it a priori} between the scaling relations $M_{0.7\;{\rm Mpc}}\sim T$ and $M_{200} \sim T^{3/2}$.} To work out a similar prediction in our case would require
solving the spherical collapse model. A precise calculation is non-trivial for two reasons: $i)$ the modified force law is not exactly $1/r^2$ on all scales and at all times; $ii)$ the collapsing matter only consists of baryons, which can dissipate energy.\footnote{The spherical collapse model has been studied in modified gravity in the context of MOND~\cite{Malekjani:2008yq}, galileons~\cite{Bellini:2012qn,Barreira:2013xea} and $f(R )$/chameleon~\cite{Martino:2008ae,Parfrey:2010uy,Brax:2010tj,Borisov:2011fu}.} To the extent that the collapse dynamics are in the $f$-regime and energy is conserved, however, then the spherical collapse calculation would proceed in the usual way, and the scaling relation~\eqref{MT} would be recovered in our model as well.\footnote{Since $f = 6 \simeq \Omega_{\rm m}/\Omega_{\rm b}$, the normalization of the mass-temperature relation would also match the standard CDM prediction.} A careful study of the spherical collapse model is beyond the scope of this paper and is left to future work.

\subsection{Profile in the central region}

\begin{figure}[t]
\centering
\includegraphics[width=3.2in]{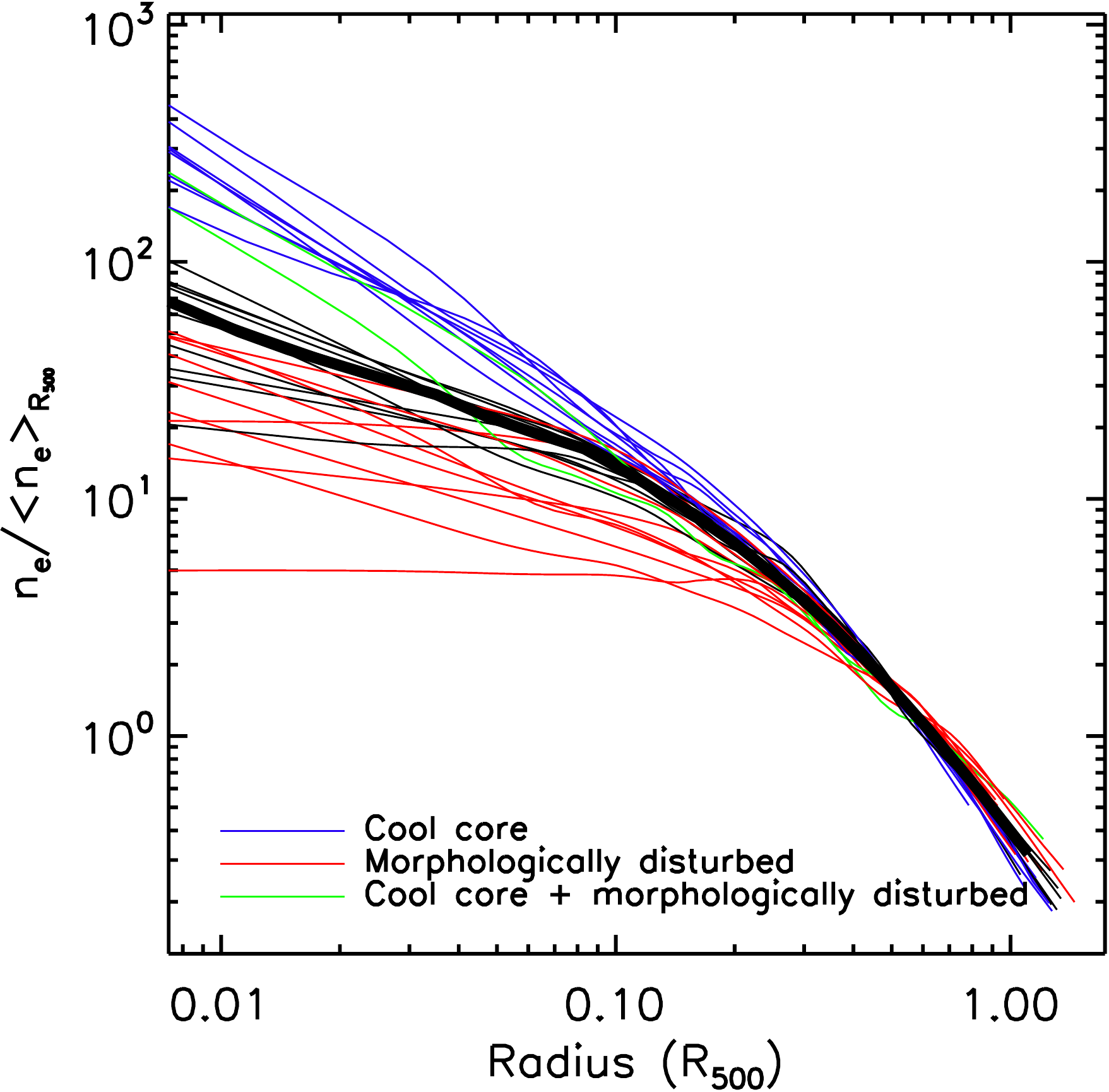}
\includegraphics[width=3.2in]{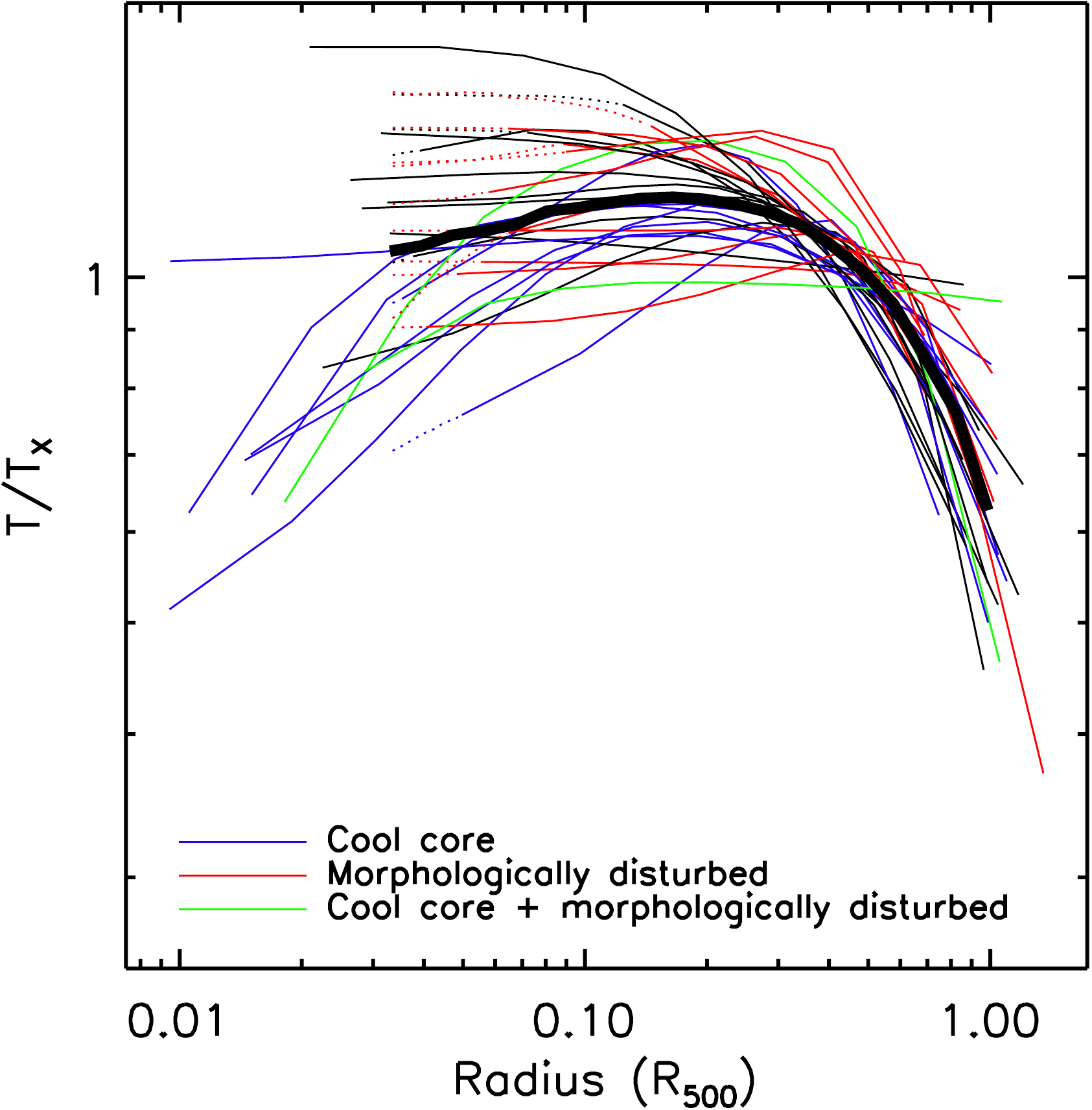}
\caption{\label{arnaudfig} Density (Left Panel) and temperature (Right Panel) profiles for the REXCESS cluster sample, reproduced from~\cite{Arnaud:2009tt}.}
\label{arnaudfigs}
\end{figure}

In the central region of the cluster, the physics is complicated by the brightest central galaxy and feedback processes. Using the REXCESS sample,~\cite{Arnaud:2009tt} obtained
a central pressure profile of the form
\be
P_{\rm central}(r ) \sim \frac{1}{r^{0.31}}\,.
\ee
Using this profile as input, the ideal gas law, our gravitational force law, and hydrostatic equilibrium, we can derive the density and temperature profiles in the central region. One subtlety is whether the central region is
in the MOND regime, or in the $f$-regime. This makes little difference, as it turns out. In either case, we find
\be
\rho_{\rm central} (r ) \sim \frac{1}{r^{1.2}}\,;\qquad ~T_{\rm central} (r ) \sim r^{0.9}\,.
\label{centralscalings}
\ee

The (scaled) density and temperature profiles of the REXCESS sample is reproduced in Fig.~\ref{arnaudfigs}. We will focus on cool-core clusters, plotted as the blue curves,
since these are relaxed clusters with minimal feedback. In the inner region ($r\lesssim 0.1~R_{500}$), the density profile becomes shallower, roughly consistent
with~\eqref{centralscalings}. The temperature of the cool-core clusters does show a drop in the inner region, though not as steep as~\eqref{centralscalings} suggests. This may be due to
feedback from the brightest central galaxy. This issue deserves closer study.

\subsection{Profile in the outer region: SZ and lensing observables}

Beyond the virial radius, we expect the density profile to become steeper than the isothermal scaling. A natural expectation is 
that the enclosed mass approaches a constant, which requires
\be
\rho(r ) \sim \frac{1}{r^{3+ \alpha}}\,;\qquad \alpha > 0\,.
\ee
In this case, the acceleration has the usual fall-off $a(r )\sim r^{-2}$, and it follows from~\eqref{hydroequiODE} that
\be
T(r ) \sim \frac{1}{r}\,.
\ee
For the pressure, this implies
\be
P(r ) \sim \rho(r )T(r )\sim \frac{1}{r^{4+ \alpha}}\,.
\label{pressure}
\ee

\begin{figure}
\centering
\includegraphics[width=3.5in]{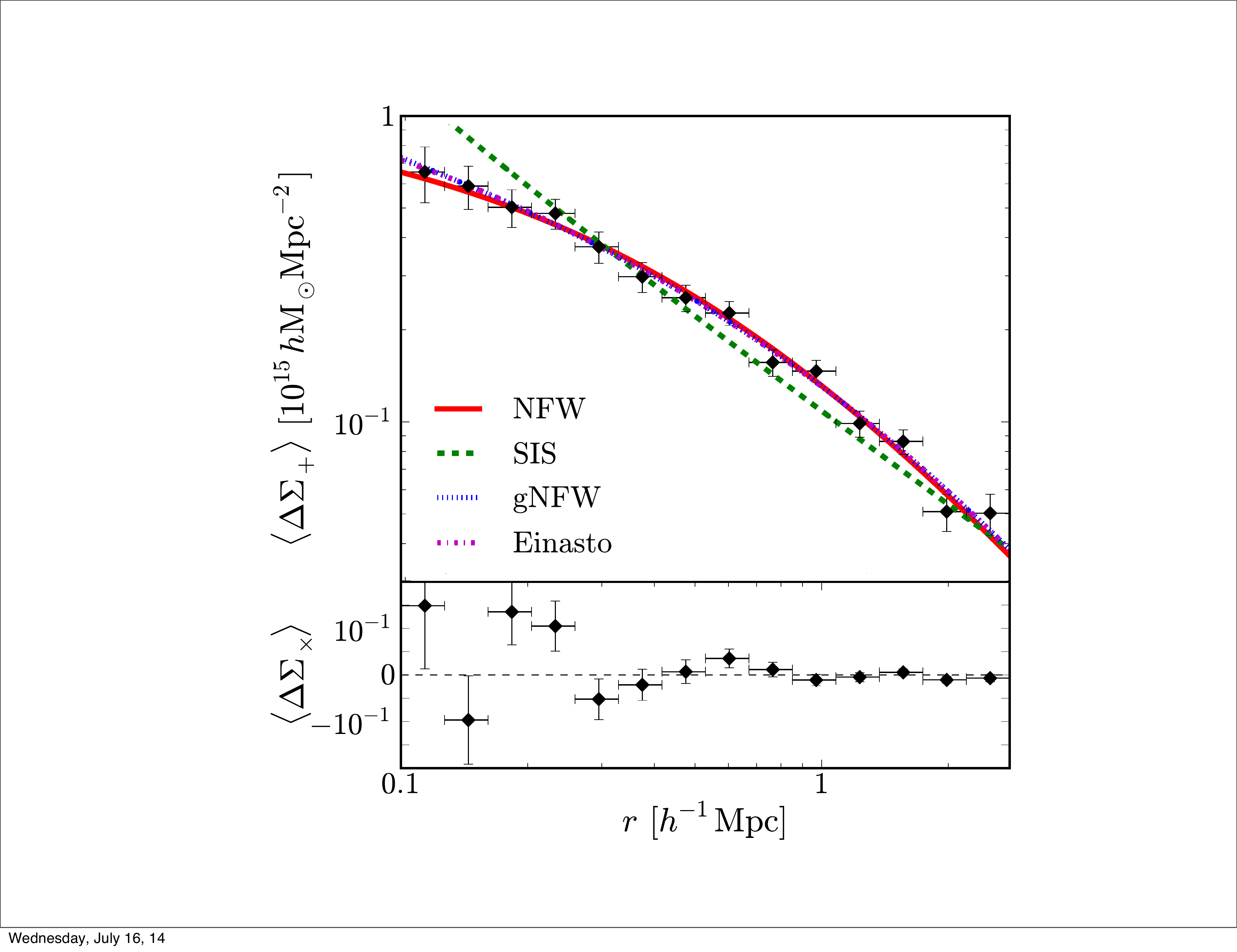}
\caption{\label{LOCUSS} \small Stacked tangential shear profile for LoCuSS clusters, reproduced from~\cite{Okabe:2013efa}.}
\end{figure}

Although X-ray measurements do not extend far enough to probe this fall-off, we can rely on SZ and weak lensing observations. The Planck satellite measured the pressure profile
for 62 nearby clusters~\cite{:2012vza}. Combining X-ray data from XMM-Newton and their own SZ data, the Planck collaboration constrained the asymptotic pressure fall-off for the stacked
sample as
\be
P(r ) \sim \frac{1}{r^{4.13}}\qquad ({\rm XMM}~\&~{\rm Planck\,\,SZ})\,,
\ee
which is consistent with~\eqref{pressure} for $\alpha \simeq 0.1$. The $T\sim 1/r$ asymptotic profile is harder to test observationally since the X-ray brightness falls off sharply with distance,
but it is consistent with the drop observed in Chandra clusters~\cite{Vikhlinin:2005mp} (see their Fig.~16) and in the REXCESS sample~\cite{Arnaud:2009tt}
(see right panel of Fig.~\ref{arnaudfigs}). Weak lensing observations extend even further, out to distances of several Mpc's. Figure~\ref{LOCUSS} (top panel), reproduced from~\cite{Okabe:2013efa}, shows the stacked
tangential shear profile for 50~massive clusters from the Local Cluster Substructure Survey (LoCuSS). For $r \;\gsim\; {\rm Mpc}$, the shear profile is steeper than the
isothermal profile $\Sigma \sim r^{-1}$ (green dashed line). The steeper slope is consistent with the NFW fall-off $\rho\sim r^{-3}$.

\subsection{Consistency check}

For consistency, we must check that clusters are in fact in the enhanced inverse-square-law regime, {\it i.e.}, 
\be
a_{\rm N} \lesssim \frac{a_0}{f^2}\,,
\label{thirdregime}
\ee
where $f\simeq 6$. Let us focus on the isothermal region, where $\rho \sim r^{-2}$. In this region, the Newtonian acceleration
is $a_{\rm N} = \Phi_{\rm N}/r$, where $\Phi_{\rm N} = {\rm constant}$. For a fiducial cluster of mass $M_{0.7\;{\rm Mpc}} = 0.8 \times 10^{13}M_\odot$
and temperature $kT = {\rm keV}$, consistent with~\eqref{mohrfit}, we obtain $\Phi_{\rm N} \simeq 6 \times 10^{-7}$. 
The inequality~\eqref{thirdregime} becomes
\be
r ~\gsim ~ \frac{2H_0}{a_0} \cdot  50~{\rm kpc}\,,
\ee
where we have used the preferred value $f =6$. The lower bound should be at most $\simeq 50~{\rm kpc}$, since clusters
are observed to be isothermal down to that scale. This requires
\be
a_0^{\rm clusters} \;\gsim\; 2 H_0\simeq 1.4\times 10^{-7}~{\rm cm}/{\rm s}^2\,.
\label{a0bestfitclusters}
\ee
In particular, had we used the MOND value inferred for galaxy fits, $a_0^{\rm galaxies} \simeq H_0/6$, we would have instead obtained a lower bound of $\simeq 600~{\rm kpc}$,
which is clearly too large.

Thus we learn that the critical acceleration $a_0$ must have some dependence on the scale or on the mass of the object.
Note that the required scale/mass-dependence is very mild --- a logarithmic dependence would do the job. If $a_0$ is related to cosmology and dark energy, as MOND proponents have been advocating for years, then it is reasonable to expect $a_0$ to approach its cosmological value $\sim H_0$ for clusters, the largest virialized objects in the universe. In the next Section, we will consider the implications
of our modified force law for smaller objects, namely galaxies and Lyman-$\alpha$ clouds.

\section{Phenomenology of Galaxies and Lyman-$\alpha$ Absorbers}
\label{galaxies}

We must ensure that our modified acceleration law does not compromise the successful MOND phenomenology for galaxies. In other words,
galaxies should lie comfortably in the intermediate regime of~\eqref{newMOND}, namely
\be
a_{\rm N} ~\gsim~ \frac{a_0^{\rm galaxies}}{f^2} = \frac{3.6\times 10^3}{f^2}~\frac{({\rm km}/{\rm s})^2}{{\rm kpc}} \,,
\label{galaxybound}
\ee
where in the last step we have assumed the standard MOND value $a_0^{\rm galaxies} = 1.2\times 10^{-8}~{\rm cm}/{\rm s}^2$. 
Below we will check~\eqref{galaxybound} individually for different classes of objects. We will find that a larger value of $f$ is required, namely $f\; \gsim\;10$, as foreseen in Table~\ref{params}.

\subsection{Spiral galaxies}

\begin{figure}
\centering
\includegraphics[width=4.5in]{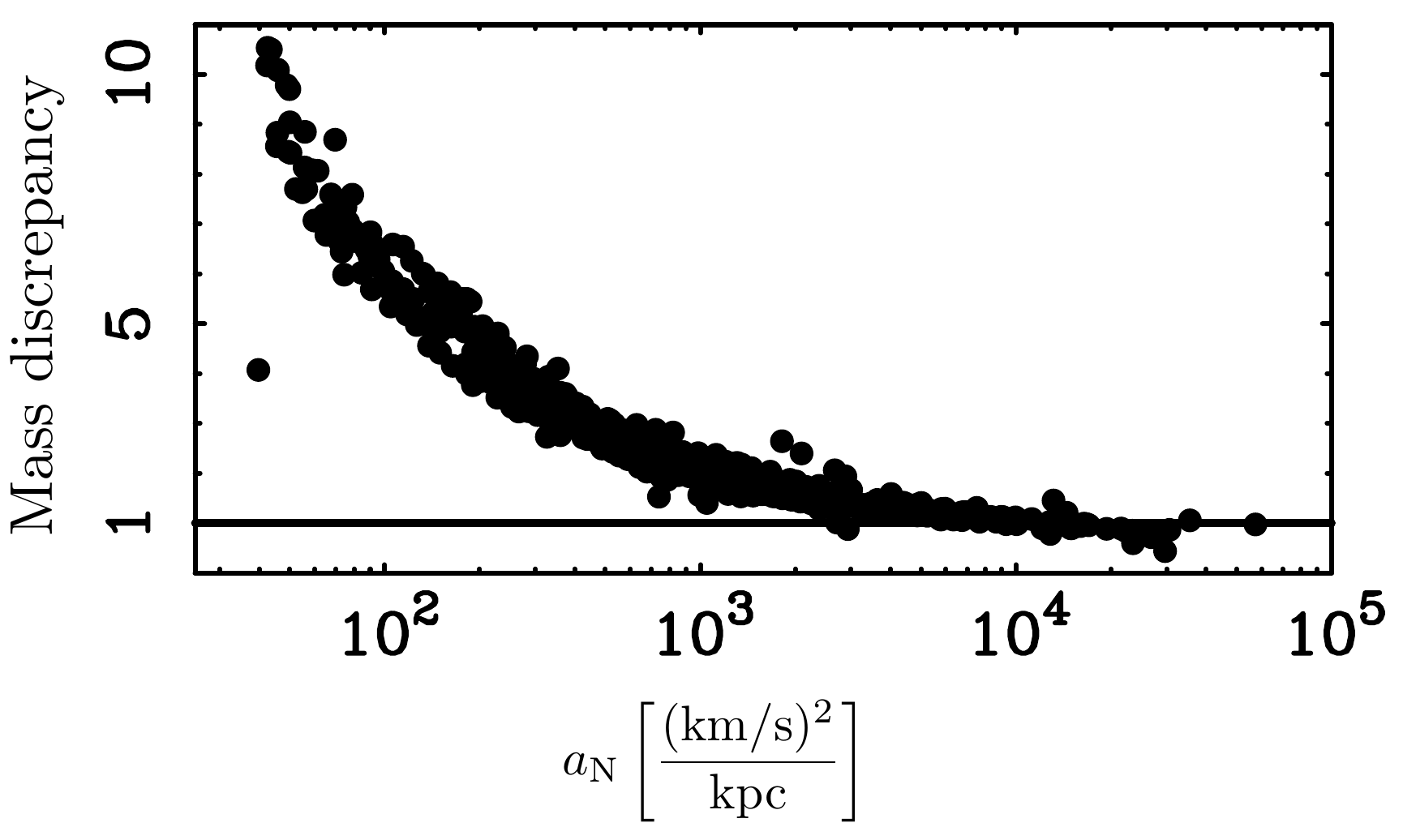}
\caption{\label{disc} \small Mass discrepancy as a function of the Newtonian acceleration for a large sample of disc galaxies, reproduced from~\cite{McGaugh:2004aw}. Each galaxy plotted has a velocity uncertainty of less than 5\%. The mass-to-light ratio was obtained using the MOND fit, as detailed in~\cite{Sanders:2002pf}.}
\end{figure}

Figure~\ref{disc}, reproduced from~\cite{McGaugh:2004aw}, shows the mass discrepancy as a function of the Newtonian acceleration for a large sample
of disc galaxies, with less than 5\% velocity uncertainties. The orbits are assumed circular. We are not concerned with the mass discrepancy, only with the range of
centripetal accelerations probed by observations. As can be read off from the plot, the smallest acceleration in the sample is $\approx 40~\frac{({\rm km}/{\rm s})^2}{{\rm kpc}}$. 
In other words,
\be
a_{\rm N} ~~\gsim~~40~\frac{({\rm km}/{\rm s})^2}{{\rm kpc}} \,.
\label{centripetaldata}
\ee
To ensure that all disc galaxies in the sample are consistent with~\eqref{galaxybound}, and therefore in the MOND regime, we must require
\be
f_{\rm galaxies} \;\gsim\; 10 \,.
\ee
(A similar bound was quoted in~\cite{Babichev:2011kq}.) Just like the critical acceleration, the $f$ parameter must also have some scale or mass dependence, ranging from $\simeq 10$ on galactic scales to $\simeq 6$ on cluster scales. 

\subsection{Ellipticals and dwarf spheroidals}

For a pressure-supported system in the MOND regime, the velocity dispersion $\sigma$ and size $R$ are related to the characteristic acceleration by
\be
a = \sqrt{a_{\rm N} a_0^{\rm galaxies}} = \frac{\sigma^2}{R}\,.
\ee
Combined with~\eqref{galaxybound}, we obtain
\be
\frac{\sigma^2}{R} ~~\gsim ~~ \frac{a_0^{\rm galaxies}}{f}\,.
\label{dispersionbound}
\ee
This is the condition for pressure-supported systems, such as elliptical and dwarf galaxies, to be in the MOND regime. 

\begin{figure}
\centering
\includegraphics[width=3.5in]{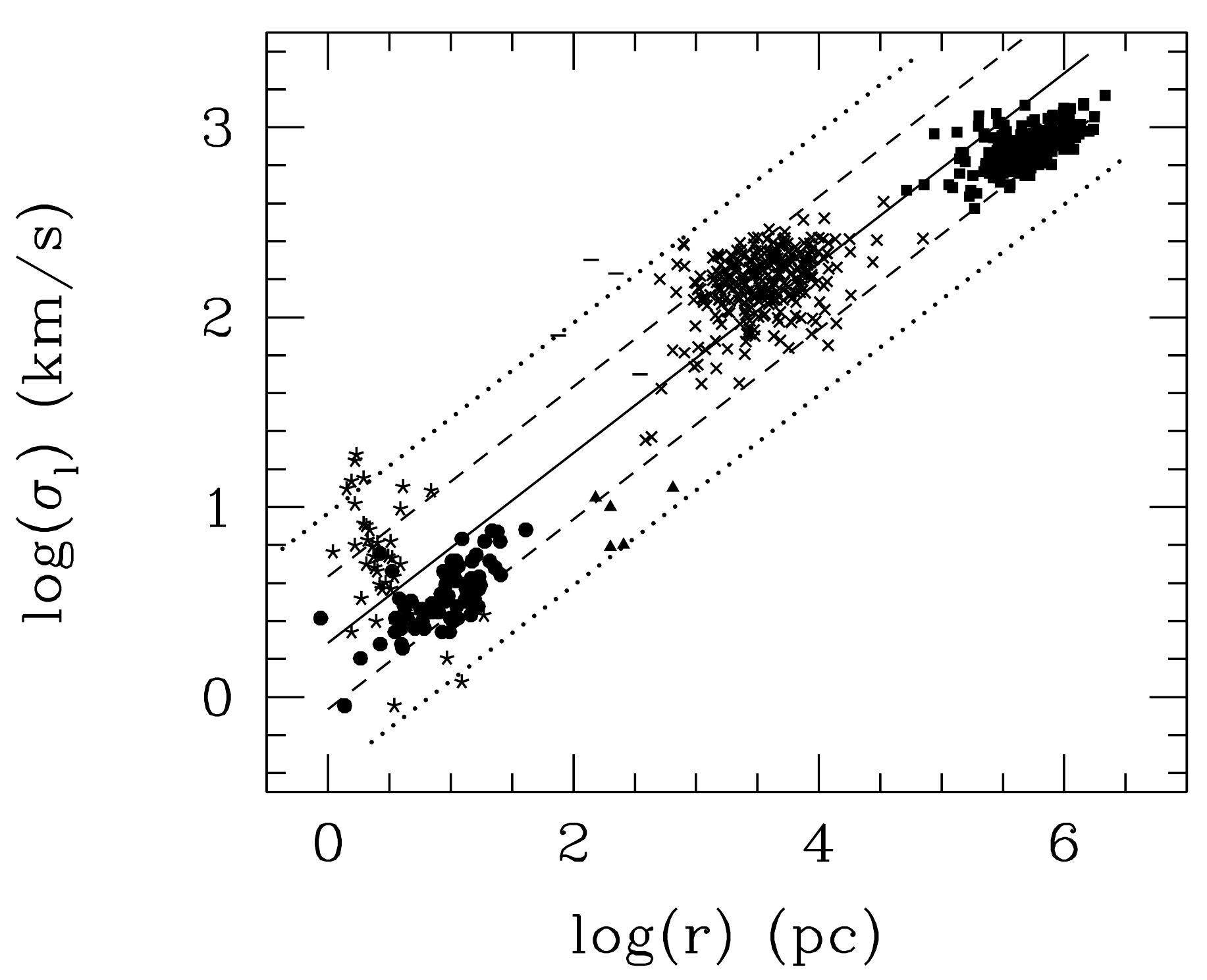}
\caption{\label{velocities} \small Line-of-sight velocity dispersion as a function of characteristic radius for pressure-supported systems, reproduced from~\cite{Sanders:2002pf}. The stars are globular clusters, the circles are massive molecular clouds, the triangles are dwarf spheroidal satellites of the Milky Way, the dashes are compact elliptical galaxies, the crosses are massive elliptical galaxies, and the squares are galaxy clusters. The solid line corresponds to $\sigma^2/r = a_0^{\rm galaxies}$. The dashed lines have slopes a factor of 5 larger or smaller than this relation. The dotted lines (added by the author) have slopes a factor of 10 larger or smaller.} 
\end{figure}

Figure~\ref{velocities}, reproduced from~\cite{Sanders:2002pf}, is a plot of $\sigma$ {\it vs} $R$ for various classes of objects, ranging from globular clusters to galaxy clusters.
The solid line corresponds to the relation $\sigma^2/R = a_0^{\rm galaxies}$; the dashed lines have a slope 5 times smaller or larger than this relation. We have added the dotted lines
to the plot with slopes 10 times larger or smaller. As inferred from the figure, nearly all objects plotted have 
\be
\frac{\sigma^2}{R} ~~\gsim ~~ 0.1 \,a_0^{\rm galaxies} \,.
\ee
This is consistent with~\eqref{dispersionbound} for $f_{\rm galaxies} \;\gsim\; 10$.

It should be mentioned that the Milky Way dwarf spheroidal satellites may pose a problem for MOND~\cite{Spergel,Milgrom:1995hz,Angus:2008vs,Hernandez:2009by,Lughausen:2014hxa}. A recent numerical analysis of the classical dwarfs (except for Ursa Minor, which appears to be out of equilibrium~\cite{Kleyna:2003zt}), carefully accounting for the external field effect, finds that MOND successfully predicts the mass-to-light ratio inferred for the most luminous dwarfs, namely Fornax and Sculptor, but underpredicts the mass-to-light ratio for Sextans, Carina and Draco~\cite{Lughausen:2014hxa}. A possible explanation within MOND is that observations have over-estimated the dynamical mass of these latter three dwarfs, for instance due to binaries or contaminant outliers, or that these systems are not in virial equilibirum. (Ultra-faint dwarfs, which would otherwise pose a grave problem for MOND, are also believed to be out of equilibrium~\cite{McGaugh:2010yr}.) On the other hand, MOND does an excellent job at explaining the observed velocity dispersions in Andromeda's dwarf satellites~\cite{McGaugh:2013wj,McGaugh:2013zqa}.

\subsection{Lyman-$\alpha$ absorbers}

Lyman-$\alpha$ clouds (which are not included in Fig.~\ref{velocities}) represent another class of pressure-supported systems. 
They are responsible for the absorption patterns in quasar spectra, also known as the Lyman-$\alpha$ forest. 
It has been argued that these systems pose a problem for MOND~\cite{Aguirre:2001fj}, as reviewed below.

The physical properties of Lyman-$\alpha$ absorbers can be derived using simple Jeans-like arguments~\cite{Schaye:2001me}. 
In the CDM framework, their estimated characteristic size is~\cite{Schaye:2001me}
\be
L \approx 1.0\times 10^2~{\rm kpc} \, \left(\frac{N_{{\rm H}_{\rm I}}}{10^{14}~{\rm cm}^{-2}}\right)^{-1/3} \left(\frac{T}{10^4~{\rm K}}\right)^{0.41} \left(\frac{\Gamma}{10^{-12}~{\rm s}^{-1}}\right)^{-1/3} \left(6 f_g\right )^{2/3}\,,
\label{lyalphasize}
\ee
where $N_{{\rm H}_{\rm I}}$ is the neutral hydrogen column density, $\Gamma$ is the hydrogen photoionization rate, and $f_g$ is the fraction of the mass in gas. The latter
is expected to be close to the cosmological value $f_g \simeq \Omega_{\rm b}/\Omega_{\rm m}\simeq 1/6$. Note that the detailed nature of dark matter plays a minor role in this derivation;
dark matter only enters through $f_g$.

The typical Newtonian acceleration in these systems is minuscule~\cite{Aguirre:2001fj},
\be
a_{\rm N}  \sim 3\times 10^{-4} a_0^{\rm galaxies} \,.
\label{lyalphaaccn}
\ee
In the MOND framework, Lyman-$\alpha$ clouds are therefore deep in the MONDian regime, with $a\simeq \sqrt{a_{\rm N} a_0}$.
However, the MOND acceleration law fails to reproduce the properties of these systems~\cite{Aguirre:2001fj}. Instead of~\eqref{lyalphasize},
MOND predicts
\be
L \approx 11~{\rm kpc} \, \left(\frac{N_{{\rm H}_{\rm I}}}{10^{14}~{\rm cm}^{-2}}\right)^{-1/5} \left(\frac{T}{10^4~{\rm K}}\right)^{0.65} \left(\frac{\Gamma}{10^{-12}~{\rm s}^{-1}}\right)^{-1/5}\qquad ({\rm MOND})\,.
\ee
This is about an order of magnitude smaller than~\eqref{lyalphasize}, and has the wrong scalings for all parameters. One caveat is the external field effect in these systems, which may restore the clouds to the desired size. See~\cite{Sanders:2002pf} for a short discussion.

In our case, the situation is different. The characteristic Newtonian acceleration~\eqref{lyalphaaccn} is so small that it violates~\eqref{galaxybound} --- Lyman-$\alpha$ clouds are
not in the MOND regime, but instead the ``$f$-regime". The relevant acceleration is an inverse-square law, like Newtonian gravity but with $G_{\rm N} \rightarrow fG_{\rm N}$. 
The analysis of~\cite{Schaye:2001me} therefore applies identically, except for the trivial replacement $f_g \rightarrow 1/f$. Specifically, instead of~\eqref{lyalphasize} we obtain
\be
L \approx 1.0\times 10^2~{\rm kpc} \, \left(\frac{N_{{\rm H}_{\rm I}}}{10^{14}~{\rm cm}^{-2}}\right)^{-1/3} \left(\frac{T}{10^4~{\rm K}}\right)^{0.41} \left(\frac{\Gamma}{10^{-12}~{\rm s}^{-1}}\right)^{-1/3} \left(\frac{6}{f}\right )^{2/3}\,.
\label{lyalphasizeourcase}
\ee
Our modified force law is nicely consistent with the observed properties of Lyman-$\alpha$ absorbers. 

It is reassuring that the $f$-regime resolves known tensions for MOND not just with one, but with two, vastly different systems: galaxy clusters and
Lyman-$\alpha$ absorbers. With clusters only, one would naturally question the justification of modifying MOND for just one class of objects. 
The fact that a simple extension like~\eqref{newMOND} is consistent with the MOND phenomenology for galaxies, while curing the known problems
of MOND in two other classes of systems, is encouraging.

\section{Problematic Observations?}

In this Section, we briefly mention two observations that may be problematic for our scenario: the Bullet Cluster and the ellipticity of dark matter halos.
Neither observable represents a show-stopper at present, but each requires closer inspection and more detailed predictions from the model.

\subsection{Bullet Cluster}

The ``Bullet'' Cluster 1E0657-57~\cite{Clowe:2003tk,Clowe:2006eq} shows lensing peaks displaced from the gas and centered around the galaxy distribution.
This is expected in CDM: the halos are made up of weakly interacting dark matter particles that fly past each other, together with the galaxies, while the baryonic plasma is
slowed down by ram pressure and ends up spatially segregated from the halos. By now observers have identified over a handful of similar merging
systems~\cite{Diego:2014eda}. 

While the Bullet system was initially hailed as ruling out MOND~\cite{Clowe:2006eq}, its asymmetric and dynamical nature makes the analysis considerably more tricky.
This issue was studied in some detail in the context of TeVeS~\cite{Angus:2006qy}. The lesson is that inferring the projected mass from weak lensing maps is subtle for such extreme asymmetric configurations
in MOND. Similarly, one would have to carefully analyze merging clusters in the context of our scenario, for instance using the relativistic theory described in Sec.~\ref{toytheory}. In particular, local gradients in $\pi$, largely ignored in our discussion so far, may be important in generating the required lensing/mass displacements.

\subsection{Dark matter halo ellipticity}

A key prediction of CDM simulations is the ellipticity of dark matter halos~\cite{Dubinski:1991bm}. This can be tested using galaxy-galaxy weak lensing observations~\cite{Hoekstra:2003pn,Mandelbaum:2005nf}. 
In MOND, however, one expects the shear signal to be approximately isotropic at large distances from the luminous matter. A detection of halo ellipticity would
therefore pose grave problems for MONDian modifications to gravity. Such a detection was claimed in~\cite{Hoekstra:2003pn}, though a subsequent analysis using
Sloan Digital Sky Survey data showed weaker ($\lesssim 2\sigma$) evidence~\cite{Mandelbaum:2005nf}. It would be very interesting to quantify the expected degree
of isotropy in our model, in particular whether $\pi$ gradients can play an important role. It may turn out that halo ellipticity can rule out the scenario.

\vspace{0.4cm}
On the flip side, the present model does better than CDM with other observables. For instance, it circumvents entirely the ``too big to fail'' problem~\cite{BoylanKolchin:2011de}
of CDM, {\it i.e.}, simulations predicting dark massive subhaloes in the Milky Way which have not been observed. The ``cusp'' problem at the core of galaxies is also obliterated
in the present framework.

\section{Conclusion}

In this paper we proposed an alternative to particle dark matter that incorporates some of the ingredients of the MOND paradigm while adding new important components.
The first new feature is a dark matter fluid, in the form of a scalar field with small equation of state and sound speed. This component is critical in reproducing the
success of CDM for the expansion history and the growth of linear perturbations. However, it does not play a major role on non-linear scales. Instead, the missing mass problem
in galaxies and clusters of galaxies is addressed via a modification of the gravitational force law.

The new force law, given by~\eqref{newMOND}, is an extension of MOND. Like MOND, the modification kicks in below some critical acceleration $a_0$. The force law is
MONDian ($a =\sqrt{a_0a_{\rm N}}$) for a while until, at very low acceleration, it reverts to an inverse-square-law with a stronger Newton's constant ($a = f a_{\rm N}$). 
The force law reduces to MOND on galactic scales and therefore piggy-backs on the empirical MONDian success at fitting galaxy rotation curves.
On cluster scales, however, the force law is in the inverse-square-law regime. We argued this explains the nearly isothermal profiles of clusters and matches the observed
temperature normalization for $f\simeq 6$. The modified force law proposed here therefore solves the well-known problems of MOND on cluster scales. By the same token, it
also successfully reproduces the features of Lyman-$\alpha$ absorbers, another problematic system for MOND~\cite{Aguirre:2001fj}.

We presented an example of a relativistic theory that realizes these features. The theory uses two scalar fields coupled in a particular way to matter. 
The first scalar is governed by the DBI action~\eqref{LDBI},
\be
{\cal L}_{\rm DBI} = -M_{\rm Pl}^2a_0^2 \sqrt{-h}\,;\qquad h_{\mu\nu} = g_{\mu\nu} + \partial_\mu\pi\partial_\nu\pi\,.
\label{Lpi}
\ee
In the limit of relativistic brane motion ($\dot{\pi}\simeq 1$, $\gamma \gg 1$), the equation of state $w = - \gamma^{-2}$ and sound speed $c_s = \gamma^{-1}$ are both small,
and the scalar behaves as dark matter. This component ensures that the CDM phenomenology on linear scales is successfully reproduced. 

The second scalar mediates the new MONDian modification of gravity. A prototypical action is the DBI-like theory~\eqref{DBIlike},
\be
{\cal L}_{\rm New\,MOND} = - \frac{(\partial\phi)^2}{f} \sqrt{ 1 + \left(\frac{2f}{3M_{\rm Pl}a_0}\right)^2 (\partial\phi)^2}\,.
\label{Lphi}
\ee
The resulting force interpolates between the MOND law for large scalar gradients ($ f|\partial\phi| \gg M_{\rm Pl}a_0$)
and an inverse-square law for small gradients ($ f|\partial\phi| \ll M_{\rm Pl}a_0$).

Ordinary matter fields are coupled to the two scalars through an effective metric~\eqref{gmat}:
\be
\tilde{g}_{\mu\nu} = e^{-2\phi/M_{\rm Pl}} h_{\mu\nu} - e^{2\phi/M_{\rm Pl}} \partial_\mu\pi \partial_\nu\pi \,.
\label{gmatconclu}
\ee
This form, inspired by the TeVeS~\cite{Bekenstein:2004ne}, is crucial for lensing mass estimates to agree with dynamical estimates. 
Unlike TeVeS, which employs a time-like vector field, our effective metric only involves scalar fields.  
A noteworthy advantage over other scalar formulations~\cite{Blanchet:2011wv} is that it is a {\it local} function of the fields. 

Many directions would be worth pursuing:

\begin{itemize}

\item The parameters $f$ and $a_0$ must be mildly scale-dependent, as summarized in Table~\ref{params}, to simultaneously fit galactic and cluster phenomenology. As mentioned in Sec.~\ref{scaledep}, 
a tantalizing possibility would be a dynamical mechanism to explain the emergence of $a_0 \sim H_0$ in the scalar Lagrangians~\eqref{Lpi} and~\eqref{Lphi}. If
$a_0$ (and $f$) can be determined cosmologically, it would be natural to expect some scale dependence as well.

\item The form of the theory, particularly the DBI action~\eqref{Lpi} and the effective metric~\eqref{gmatconclu}, strongly suggests a geometric interpretation in terms of
branes moving in extra-dimensional bulk space-times. A geometric realization, if possible, might point the way towards a string theory embedding. Even at the level
of effective field theory, a geometric embedding can unveil new symmetries, inherited from bulk isometries, whose 4d realization is highly non-trivial. 

\item Detailed predictions should be worked out for the Bullet Cluster and similar merging systems. This will require a careful modeling of the $\pi$ and $\phi$ profiles in time-dependent,
asymmetric configurations.

\end{itemize}

\noindent {\bf Acknowledgements:} We thank Anthony Aguirre, Andy Albrecht,  Lasha Berezhiani, Clare Burrage, Paolo Creminelli, Jose Maria Diego, Benjamin Elder, Elise Jennings, Austin Joyce, Stacy McGaugh, David Spergel, Paul Steinhardt, Mark Trodden, Junpu Wang and Matias Zaldarriaga for useful discussions. We are especially grateful to Nima Arkani-Hamed and Erik Verlinde for inspiring discussions, and to Bhuvnesh Jain and Ravi Sheth for many helpful conversations about observations. We also thank an anonymous referee for very helpful suggestions. This work was supported in part by NSF CAREER Award PHY-1145525 and NASA ATP grant NNX11AI95G.

\section*{Appendix: DBI as Dark Matter and Dark Energy}
\renewcommand{\theequation}{A-\Roman{equation}}
\setcounter{equation}{0} 

\begin{figure}[t]
\centering
\includegraphics[width=3.5in]{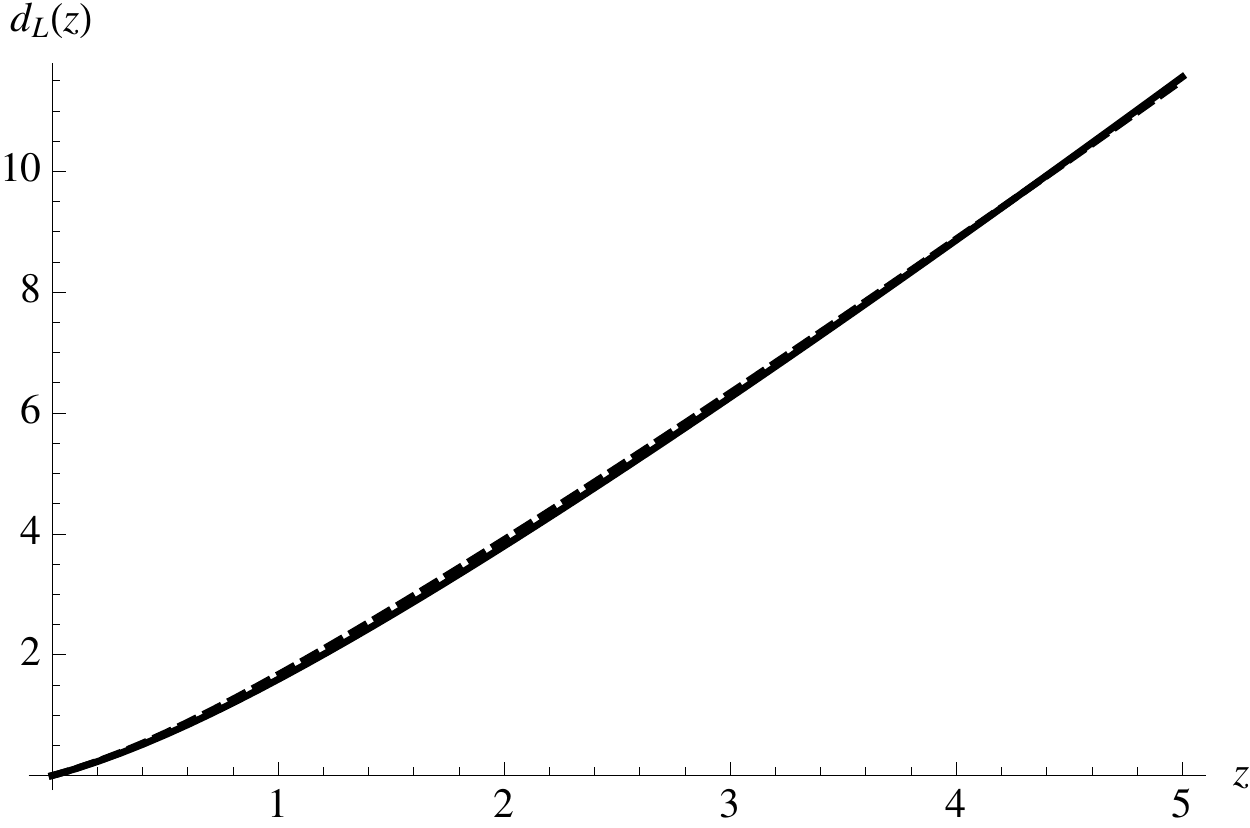}
\caption{\label{lumin} \small Luminosity distance as a function of redshift for $\Lambda$CDM with $\Omega_{\rm m} = 0.25$ (solid line) and DBI with $C = 0.32$. The percentage difference is less than 5\% over the entire redshift range.}
\end{figure}

A virtue of the DBI scalar is that it can act both as dark matter and as dark energy. Let us restore the arbitrary scale $M$
and write the DBI action~\eqref{LDBI} as
\be
{\cal L}_\pi = -M^4\sqrt{1-X}\,.
\ee
As shown in Sec.~\ref{DBIscalar}, the energy density is given by
\be
\rho = M^4 \gamma = M^4 \sqrt{1 + \frac{C^2}{a^6}}\,.
\ee
It behaves as dust at early times and as a cosmological constant at late times. The expansion history is somewhat different, and possibly distinguishable, from $\Lambda$CDM. 
The expansion rates in the two cases are given by, assuming a spatially flat ($k=0$) universe,
\bea
\nonumber
\frac{H(z)}{H_0} &=& 1 - \Omega_{\rm m} + \frac{\Omega_{\rm m}}{a^3} \qquad ~~~~~~~~~~(\Lambda{\rm CDM}) \\
\frac{H(z)}{H_0} &=& \left(\frac{1 + C^2(1 + z)^6}{1 + C^2} \right)^{1/4} \qquad ({\rm DBI})\,,
\eea
where in the DBI case we wrote $M^4 = \frac{3H_0^2M_{\rm Pl}^2}{\sqrt{1+C^2}}$. The luminosity distances are shown in Fig.~\ref{lumin} as functions of redshift. The solid curve is the
$\Lambda$CDM luminosity distance with $\Omega_{\rm m} = 0.25$; the dashed curve, fitted by eye, is the DBI distance with $C = 0.32$. The percentage difference is less than 5\%
over the entire redshift range, peaking at $z\simeq 1$. 

The main drawback of this scenario is that the coupling to matter must be modified in a non-local way. Indeed, since $\dot{\pi}\neq 1$ at the present time ({\it e.g.}, $\dot{\pi}\simeq 0.3$ for $C = 0.32$),
the effective metric coupling to matter will not be of the form~\eqref{lensingform} required for lensing. One can instead couple matter to the following metric
\be
\tilde{g}_{\mu\nu} = e^{-2\phi/M_{\rm Pl}} g_{\mu\nu} - 2u_\mu u_\nu  \sinh \frac{2\phi}{M_{\rm Pl}} \,,
\label{gmatapp}
\ee
where $u_\mu \equiv \partial_\mu\pi/\sqrt{X}$ is the unit time-like vector for the DM fluid. At the linear level, this does reduce to the form~\eqref{lensingform} required to match lensing observables.
However, the effective metric now depends on $\pi$ in a non-local way. This is why we focused on the `pure-dust' behavior in the main text, since it allows the local effective metric~\eqref{gmat}.
Nevertheless, the connection to dark energy is tantalizing, and it would be interesting to further explore this version of the scenario.

\end{document}